\documentclass[aps,onecolumn,preprintnumbers,nofootinbib,superscriptaddress]{revtex4}
\usepackage{standalone}
\usepackage{tikz}
\usepackage{amsmath}
\usepackage{graphicx}
\usepackage{amsfonts}
\usepackage{array}
\usepackage{amsthm}
\usepackage{bm}
\usepackage{palatino}
\usepackage{mathpazo}
\usepackage{supertabular}
\usepackage{subfig}
\usepackage[breaklinks]{hyperref}
\usepackage{color}
\usepackage{caption}
\usepackage[font=small,labelfont=bf,justification=justified,format=plain]{caption}
\usepackage[font=small,labelfont=bf,justification=justified]{caption}
\usepackage[labelfont=bf,labelsep=quad,justification=justified]{caption}
\usepackage{epstopdf}
\newcommand{\be}{\begin{equation}}
\newcommand{\ee}{\end{equation}}
\newcommand{\ba}{\begin{eqnarray}}
\newcommand{\ea}{\end{eqnarray}}
\newcommand{\bal}{\begin{align}}
\newcommand{\eal}{\end{align}}

\newcommand{\bw}{\begin{widetext}}
\newcommand{\ew}{\end{widetext}}

\providecommand{\keywords}[1]
{
	\small	
	\textbf{\textit{Keywords:}} #1
}
 \allowdisplaybreaks
\begin{document}

\title{Lyapunov exponents and geodesic stability of Schwarzschild black hole in the non-commutative gauge theory of gravity}

\author{Abdellah Touati}
\email{touati.abph@gmail.com}
\affiliation{Department of Physics, Faculty of Sciences of Matter, University of Batna-1, Batna 05000, Algeria}

\author{Slimane Zaim\footnote{Corresponding author.}}
\email{zaim69slimane@yahoo.com}
\affiliation{Department of Physics, Faculty of Sciences of Matter, University of Batna-1, Batna 05000, Algeria}

\begin{abstract}
In this paper, we study the stability of geodesic motion for both massive and massless particles using Lyapunov exponents in the non-commutative (NC) Schwarzschild black hole (BH) via the gauge theory of gravity. As a first step, we investigate both time-like and null radial motion of particles. The main result in NC geometry shows that the particles take an infinite proper time to reach the NC singularity (infinite affine parameter time framework for photons). The proper time and coordinate time of Lyapunov exponents, as well as their ratio for time-like geodesics in circular motion around this black hole, exhibit a new behavior, revealing a new range of stable circular orbits between unstable ones. Next, we analyze the circular motion of photons, where the results indicate the presence of a new photon sphere near the event horizon, which is not allowed in the commutative case. The Lyapunov exponent in this geometry confirms the instability of the outer photon sphere and the stability of the inner one.  Moreover, we study the effect of non-commutativity on the black hole shadow radius and find a similarity between non-commutativity and the mass of a black hole. Finally, using experimental data from the Event Horizon Telescope, we show that the non-commutativity parameter is of the order of $\Theta^{\text{Phy}} \sim 10^{-34} \, m$.  

\end{abstract}


\date{\today}
\maketitle

\section{Introduction}

The investigation of geodesic motion for test particles around compact celestial bodies is fundamental in astrophysics, offering valuable insights into the nature of neutron stars, black holes (BHs), and other dense astrophysical objects. The trajectory of particles moving around a BH encodes information about the underlying spacetime geometry, which is shaped by the strong gravitational field of the compact object. This study focuses on analyzing how non-commutative (NC) geometry influences the geodesic motion of test particles near a Schwarzschild BH, employing the Lyapunov exponent \cite{Lyapunov5} as a diagnostic tool. Two key types of geodesic motion often examined in BH physics are radial and circular trajectories. A wide range of studies has previously explored geodesic motion around BHs \cite{pradhan,villanueva,pugl1,pugl2,Geodesic1.1,Geodesic3.1,Geodesic5.3}. The classification of stable and unstable orbits sheds light on the spacetime geometry surrounding these objects. The Lyapunov exponent $\lambda$ provides a means to evaluate the stability of equatorial circular geodesics in BH spacetimes. Depending on the nature of the orbits whether stable, unstable, or marginally stable the Lyapunov exponent helps in determining their properties. Circular geodesics in a Schwarzschild BH are integrable, meaning they exhibit no chaotic behavior. In particular, unstable orbits have a positive Lyapunov exponent \cite{Lyapunov5}, while stable and marginally stable orbits correspond to imaginary and zero values of the exponent, respectively \cite{Lyapunov1,Lyapunov5}. The importance of geodesic stability has driven numerous studies that utilize Lyapunov exponents to explore various BH spacetimes \cite{geostab12,geostab11,geostab3,geostab9,geostab4,geostab1,geostab2,geostab7}.

Another critical feature of geodesic motion around BH spacetimes is the photon sphere, a region where photons are forced into circular orbits due to the strong gravitational field. This region plays a significant role in understanding the behavior of light near compact objects and is closely tied to the study of unstable circular orbits. The photon sphere also has implications for observables such as the BH shadow \cite{ehc}, which is influenced by the dynamics of photons in this region \cite{geostab3,photonsphere1,photonsphere2,photonsphere4,radial1,photonspheres,BHShadow1}.

Recent research has investigated how non-commutativity alters geodesic dynamics using multiple approaches, such as modeling the NC matter distribution through Gaussian functions \cite{kuniyal,iftikhar,rahaman,NCGD1,NCGD3} and Lorentzian profiles \cite{NCLD2,NCLD3,NCLD1}, in addition to the star product formalism, as well as Bopp’s transformation \cite{ulh1,larbi1} and NC gauge gravity \cite{abdellah1,abdellah4,NCGTG2,NCGTG3,NCGTGmassdeformed1,NCGTGmassdeformed2,NCGTGmassdeformed3}. In previous work \cite{abdellah1}, we identified a new class of particle motion near the event horizon in an NC spacetime for massive test particles. Here, we extend this analysis to massless particles and other motion types, providing a more comprehensive study of geodesic stability. In addition, in Ref.\cite{Tajron1}, the authors showed the existence of a new correction for the tetrad fields in the gravitational gauge theory, which was not present in the old model \cite{cham1}. We have taken this new correction into account in our study. This work contributes to the ongoing exploration of orbital stability using Lyapunov exponents in NC spacetimes \cite{geostab14,geostab13,geostab5}.

The main objective of this study is to explore the stability of geodesic motion in NC spacetime by analyzing the Lyapunov exponent for different types of particles. We aim to understand how quantum gravity effects, introduced by NC geometry, impact orbital behavior. This investigation enhances our knowledge of stability and chaos in deformed spacetimes, with potential implications for observables such as the innermost stable circular orbit and the BH shadow. NC geometry, which is motivated by string theory \cite{seiberg1}, is based on the quantization of spacetime, leading to the discretization of gravitational interactions. These quantum gravitational corrections become particularly relevant in the strong-field regions surrounding BHs and should be considered in such analyses.

In non‑commutative geometry, the position operators obey the relation:
\begin{equation}
	[x^{\mu},x^{\nu}]_* = i\Theta^{\mu\nu},
\end{equation}
where $\Theta^{\mu\nu}$ an antisymmetric matrix that establishes a fundamental discretization scale for spacetime. This modification affects the Heisenberg uncertainty principle, leading to a minimum measurable length on the order of the Planck scale:
\begin{equation}
	\Delta x^{\mu}\Delta x^{\nu} \geq \frac{1}{2} |\Theta^{\mu\nu}|.
\end{equation}
In this geometry, the usual function multiplication is replaced by the Moyal star product "$*$", which is defined for two functions $f(x)$ and $g(x)$ as follows:
\begin{align}\label{eqt2.25}
	(f*g)(x) &= f(x) e^{\frac{i}{2} \Theta^{\mu\nu} \overleftarrow{\partial_{\mu}} \overrightarrow{\partial_{\nu}}} g(x) \\
	&= f(x) g(x) + \frac{i}{2} \Theta^{\mu\nu} \partial_{\mu} f(x) \partial_{\nu} g(x) - \frac{1}{8} \Theta^{\mu\nu} \Theta^{\rho\eta} \partial_{\mu} \partial_{\rho} f(x) \partial_{\nu} \partial_{\eta} g(x) + \mathcal{O}(\Theta^3).
\end{align}

In this study, we adopt the NC gauge gravity approach in conjunction with the Seiberg-Witten (SW) map \cite{seiberg1} and star products. NC gauge gravity is a candidate extension of general relativity, maintaining action invariance under NC transformations via the SW map \cite{action1,action3}. We utilize this framework to derive the NC Schwarzschild metric, then compute the NC corrections to the effective potential and analyze geodesic motion for different particle types. Our findings suggest that both massive and massless particles are prevented from reaching the NC singularity within a finite proper time or affine parameter while also revealing the presence of novel stable circular orbits near the event horizon distinct from those predicted in other NC models. Furthermore, we assess the stability of these orbits through Lyapunov exponent analysis, confirming their stability in NC spacetime. Additionally, we determine NC corrections to the BH shadow radius and use observational data to estimate the NC parameter through different phenomena, yielding a fundamental length close to the Planck scale.

The structure of this paper is as follows: Section \ref{Sec:NCSBH} overview the NC corrections to the metric field using the star product and SW maps. Section \ref{Sec:GSNCSBH} derives the NC correction to both the geodesic equations and the effective potential, incorporating second-order corrections in $\Theta$. We examine radial and circular motion for both massive and massless particles, studying orbital stability through the Lyapunov exponent. This is followed by an exploration of NC corrections to the BH shadow radius and studying the limit of the NC parameter using observational data for different phenomena. The final section summarizes our key results and conclusions.


\section{Non-commutative corrections for Schwarzschild metric}\label{Sec:NCSBH}

To construct the NC metric for the Schwarzschild BH we follow the methodology outlined in our previous works \cite{abdellah1,abdellah6,abdellah2}, and employ both the Seiberg-Witten (SW) map and the star ($\ast$)-product. To obtain the NC gauge
potential of gravity, represented by the tetrad fields, we utilize the
perturbation expansion of the SW map for the tetrad fields $\hat{e}^{a}_{\mu}(x,\Theta)$. This expansion is carried out as a power series in $\Theta$ up to the second order, as derived in Ref. \cite{Tajron1}.
\begin{equation}
	\hat{e}^{a}_{\mu}(x,\Theta)=e^{a}_{\mu}(x)-i\Theta^{\nu\rho}e^{a}_{\mu\nu\rho}(x)+\Theta^{\nu\rho}\Theta^{\lambda\tau}e^{a}_{\mu\nu\rho\lambda\tau}(x)+\mathcal{O}(\Theta^{3})\label{eq:SWM}
\end{equation}
where the first and second order corrections are given respectively by 
\begin{align}
	e^{a}_{\mu\nu\rho}(x)=\,\frac{1}{4}[\omega^{ac}_{\nu}\partial_{\rho}e^{d}_{\mu}+(\partial_{\rho}\omega^{ac}_{\mu}+F^{ac}_{\rho\mu})e^{d}_{\nu}]\eta_{cd}
\end{align}
and
\begin{align}
	e^{a}_{\mu\nu\rho\lambda\tau}&=\frac{1}{16}\Bigg[\left(2\{F_{\tau\nu},F_{\mu\rho}\}^{ab}e^{c}_{\lambda}-\omega^{ab}_{\lambda}(D_{\rho}F_{\tau\mu}^{cd}+\partial_{\rho}F_{\tau\mu}^{cd})e^{m}_{\nu}\eta_{dm}-\{\omega_{\nu},(D_{\rho}F_{\tau\mu}+\partial_{\rho}F_{\tau\mu})\}^{ab}e^{c}_{\lambda}\right.\notag\\
	&\left.-\partial_{\tau}\{\omega_{\nu},(\partial_{\rho}\omega_{\mu}+F_{\rho\mu})\}^{ab}e^{c}_{\lambda}-\omega^{ab}_{\lambda}\left(\omega^{cd}_{\nu}\partial_{\rho}e^{m}_{\mu}+\left(\partial_{\rho}\omega_{\mu}^{cd}+F_{\rho\mu}^{cd}\right)e^{m}_{\nu}\right)\eta_{dm}+2\partial_{\nu}\omega_{\lambda}^{ab}\partial_{\rho}\partial_{\tau}e^{c}_{\mu}\right.\notag\\ 
	&\left.-2\partial_{\rho}\left(\partial_{\tau}\omega_{\mu}^{ab}+F_{\tau\mu}^{ab}\right)\partial_{\nu}e^{c}_{\lambda}-\{\omega_{\nu},(\partial_{\rho}\omega_{\lambda}+F_{\rho\lambda})\}^{ab}\partial_{\tau}e^{c}_{\mu}-\left(\partial_{\tau}\omega_{\mu}^{ab}+F^{ab}_{\tau\mu}\right)\left(\omega^{cd}_{\nu}\partial_{\rho}e^{m}_{\lambda}\right.\right.\notag\\
	&\left.\left.+\left((\partial_{\rho}\omega^{cd}_{\lambda}+F^{cd}_{\rho\lambda})\right)e^{m}_{\nu}\right)\eta_{dm}\right)\eta_{cb}-\omega^{ac}_\lambda\omega^{db}_\nu e^f_\rho R^{gm}_{\tau\mu}\eta_{cd}\eta_{fg}\eta_{bm}\Bigg],\label{eq:}        
\end{align}
where
\begin{align}
	\{\alpha,\beta\}^{ab}=\left(\alpha^{ac}\beta^{db}+\beta^{ac}\alpha^{db}\right)\eta_{cd},\quad &[\alpha,\beta]^{ab}=\left(\alpha^{ac}\beta^{db}-\beta^{ac}\alpha^{db}\right)\eta_{cd}\\
	D_{\mu}F_{\rho\sigma}^{ab}=\partial_{\mu}F^{ab}_{\rho\sigma}+&\left(\omega_{\mu}^{ac}F^{db}_{\rho\sigma}-\omega_{\mu}^{db}F^{ac}_{\rho\sigma}\right)\eta_{cd}.
\end{align}
where the commutative gauge gravity potentials are denoted by $e_{a}^{\mu }$ and $\omega^{ab}_{\mu}$, which are the tetrad fields and the spin connection respectively.

Then a symmetric real metric component $\hat{g}_{\mu \nu }$ can be
constructed according to \cite{chai1}:
\begin{equation}\label{eq:22}
	\hat{g}_{\mu \nu }=\frac{1}{2}(\hat{e}_{\mu }^{a}\ast \hat{e}_{\nu a}^\dagger+\hat{e}_{\nu }^{a}\ast \hat{e}_{\mu a}^\dagger)\,.
\end{equation}
where $\hat{e}_{\nu a}^{\dagger}$  represents the complex conjugate of the deformed tetrad fildes fields $\hat{e}_{\nu a}$.

In space-space non-commutativity, we choose the coordinate system so that the NC matrix $\Theta^{\mu\nu}$ has the following form
\begin{equation}
	\Theta^{\mu\nu}=\left(\begin{matrix}
		0	& 0 & 0 & 0 \\
		0	& 0 & 0 & \Theta \\
		0	& 0 & 0 & 0 \\
		0	& -\Theta & 0 & 0
	\end{matrix}
	\right), \qquad \mu,\nu=0,1,2,3\label{eqt2.34}
\end{equation}
In order to obtain the NC correction to the metric, we follow the same steps as in Ref.~\cite{abdellah1} and adopt the same choice of the general tetrad fields:
\begin{equation}\label{eq:tetrad}
	e^{a}_{\mu}=\left(\begin{array}{cccc}
		\left(1-\frac{2 m}{r}\right)^{\frac{1}{2}} & 0 & 0 & 0 \\ 
		0 & \left(1-\frac{2 m}{r}\right)^{-\frac{1}{2}}\sin\theta \cos\phi & r \cos\theta \cos\phi & -r \sin\theta \sin\phi \\ 
		0 & \left(1-\frac{2 m}{r}\right)^{-\frac{1}{2}}\sin\theta \sin\phi & r \cos\theta \sin\phi & r \sin\theta \cos\phi \\ 
		0 & \left(1-\frac{2 m}{r}\right)^{-\frac{1}{2}}\cos\theta & -r \sin\theta & 0
	\end{array} \right),
\end{equation}
Using the definition of Equation \eqref{eq:22}, together with the NC quadrupole fields, we obtain the non-zero components up to the second order of the NC parameter $\Theta$ for the deformed metric $\hat{g}_{\mu \nu}$ in the equatorial plane $\theta = \pi/2$. The NC line element of this new deformed spacetime can be written as follows:
\begin{align}\label{eq:4.1}
	d\hat{s}^2 = \hat{g}_{tt}(r,\Theta)c^{2}dt^{2} + \hat{g}_{rr}(r,\Theta)dr^{2} + \hat{g}_{\phi\phi}(r,\Theta)d\phi^{2}.
\end{align}
where the NC metric component are given by $\hat{g}_{\mu \nu}$: 
\begin{subequations}
\begin{align}
	-\hat{g}_{tt}=&\left(1-\frac{2 m}{r}\right)+\left\{\frac{m\left(88m^2+mr\left(-77+15\sqrt{1-\frac{2m}{r}}\right)-8r^2\left(-2+\sqrt{1-\frac{2m}{r}}\right)\right)}{8 r^4(-2m+r)}\right\}\Theta^{2}+\mathcal{O}(\Theta^{4})\label{eq:2.13}\\
	\hat{g}_{rr}=&\left(1-\frac{2 m}{r}\right)^{-1}+\left\{\frac{m \left(-48 m^2+55 m r+4 r^2\right)-4 m^2 r \sqrt{1-\frac{2 m}{r}}+4 m r^2 \sqrt{1-\frac{2 m}{r}}}{16 r^2 (r-2 m)^3}\right\}\Theta^{2}+\mathcal{O}(\Theta^{4})\label{eq:2.13'}\\
	\hat{g}_{\theta\theta}=&r^{2}+\left\{\frac{m\left(m\left(10-6\sqrt{1-\frac{2m}{r}}\right)-\frac{8m^2}{r}+r\left(-3+5\sqrt{1-\frac{2m}{r}}\right)\right)}{8(-2m+r)^2}\right\}\Theta^{2}+\mathcal{O}(\Theta^{4})\label{eq:2.13''}\\
	\hat{g}_{\phi\phi}=&r^{2}+\left\{\frac{5}{8}-\frac{3}{8}\sqrt{1-\frac{2m}{r}}-\frac{2m}{r}+\frac{m\sqrt{1-\frac{2m}{r}}(-2m+7r)}{8(-2m+r)^2}\right\}\Theta^{2}+\mathcal{O}(\Theta^{4})\label{eq:2.13'''}
\end{align}
\end{subequations}
For a spherically symmetric Schwarzschild black hole in NC coordinates, the metric (\ref{eq:2.13}-\ref{eq:2.13'''}) exhibits anisotropy, which is physically reasonable. This anisotropy arises because noncommutativity effectively acts as a mass density within the BH, leading to an asymmetric distribution of this density along the coordinate axes. This asymmetry is a consequence of the deformed geometry of space, further reinforcing the anisotropic nature of the Schwarzschild metric in this context. Notably, two limiting cases can be observed: as $r$ approaches infinity, the metric (\ref{eq:2.13}-\ref{eq:2.13'''}) reduces to the Minkowski solution, and when the noncommutativity parameter $\Theta$ approaches zero, the metric (\ref{eq:2.13}-\ref{eq:2.13'''}) reverts to the standard Schwarzschild solution in the commutative case. Moreover, this geometry shifts the singularity at $r=0$ to a new finite radius, $r^{NC}_{\text{singularity}}=2m$ \cite{abdellah2}.

Interestingly, Eqs.~(\ref{eq:2.13}-\ref{eq:2.13'''}) reveal that $\hat{g}_{tt}\hat{g}_{rr} \neq 1$, which is a necessary condition for satisfying all four energy conditions: the null energy condition, weak energy condition, dominant energy condition, and strong energy condition. This implies that the NC BH exhibits double photon spheres (for further details, see \cite{photonspheres}).

As observed, this metric exhibits two distinct surfaces, corresponding to the static limit surface and the event horizon, respectively. These surfaces are determined by solving the following equations:  
\begin{equation}
	\hat{g}_{00} = 0 \, , \quad \frac{1}{\hat{g}_{rr}} = 0.
\end{equation}
The solutions to these equations give us, respectively, the NC static limit surface and the event horizon of the NC Schwarzschild BH \cite{abdellah1,abdellah2}.
\begin{equation}
	r_{sls}^{\mathrm{NC}}=r_{h}\left[1+\left(\frac{\Theta}{r_h}\right)\left(\frac{4\sqrt{5}+1}{16\sqrt{10}}\right)+\left(\frac{\Theta}{r_h}\right)^2\left(\frac{10+\sqrt{5}}{64}\right)\right],\label{eq:2.14'}
\end{equation}
and 
\begin{equation}
	r_{h}^{NC}=r_{h}\left[ 1+\frac{1}{4}\left( \frac{\Theta }{r_{h}}\right)^{2} \right].\label{eq:2.14}
\end{equation}
Our results differ from those reported in Ref. \cite{Tajron1}, due to the solution of the NC event horizon equation ($1/\hat{g}_{rr} = 0$) up to the second order in $\Theta$, and different choices of tetrads fields.
It is clear that in the limit of $\Theta =0$, we recover the commutative event horizon $r_{h}=2m$.


\section{Geodetic motion and stability of orbits around a non-commutative Schwarzschild black hole}\label{Sec:GSNCSBH}

In this section, we analyze the stability of circular orbits constrained to the equatorial plane (i.e., $\theta = \pi/2$) within the framework of the NC Schwarzschild spacetime. In this geometry \eqref{eq:4.1}, the motion of a test particle is governed by the following Lagrangian:
\begin{align}
	2\mathcal{L} = \hat{g}_{tt}(r,\Theta)c^{2}\dot{t}^{2} + \hat{g}_{rr}(r,\Theta)\dot{r}^{2} + \hat{g}_{\phi\phi}(r,\Theta)\dot{\phi}^{2}, \label{eqt3.49}
\end{align}
where the dot denotes differentiation with respect to the affine parameter $\lambda$.

Since the Lagrangian does not explicitly depend on $t$ and $\phi$, two conserved quantities arise along the path of the test particle, which correspond to the energy ($E$) and angular momentum ($l$) respectively. Consequently, the expressions for $\dot{t}$ and $\dot{\phi}$ are given by:
\begin{align}
	\dot{t} = \frac{E}{c^{2}\hat{g}_{tt}(r,\Theta)}, \quad 
	\dot{\phi} = \frac{l}{\hat{g}_{\phi\phi}(r,\Theta)}. \label{eq:4.5}
\end{align}

Substituting these relations into the spacetime invariant equation: $$\hat{g}_{tt}U^{t}U^{t}+\hat{g}_{rr}U^{r}U^{r}+\hat{g}_{\phi\phi}U^{\phi}U^{\phi} \equiv -h$$
where $U^{\mu} = c^{-1}\frac{dx^{\mu}}{d\tau}$ is the four-velocity, and with some algebra we obtain the radial equation:
\begin{equation}
	\dot{r}^{2} = -\frac{E^{2}}{c^{2}\hat{g}_{tt}(r,\Theta)\hat{g}_{rr}(r,\Theta)} - \frac{1}{\hat{g}_{rr}(r,\Theta)}\left(\frac{l^{2}}{\hat{g}_{\phi\phi}(r,\Theta)} + hc^{2}\right), \label{eq:4.6}
\end{equation}
where $h = 1$ for massive particles and $h = 0$ for massless particles.

By incorporating the specific metric components, the radial equation simplifies to:
\begin{equation}
	\dot{r}^{2} + V_{\text{eff}}(r,\Theta) = 0, \label{eqt 3.68}
\end{equation}
where the effective potential $V_{\text{eff}}(r,\Theta)$ for a test particle in NC spacetime up to second order in the NC parameter $\mathbf{\Theta}$, is given by \cite{abdellah1}:
\begin{align}
	V_{\text{eff}}(r,\Theta)&=\left(1-\frac{2 m}{r}\right)\left(\frac{l^{2}}{r^{2}}+hc^{2}\right)-\frac{E^{2}}{c^2}+\Theta^{2}\left\{-\frac{l^{2}}{r^{4}}\left(1-\frac{2m}{r}\right)\left(\frac{5}{8}-\frac{3}{8}\sqrt{1-\frac{2m}{r}}+\frac{m\sqrt{1-\frac{2m}{r}}(-2m+7r)}{8(-2m+r)^2}\right.\right.\notag\\
	&\left.\left.-\frac{2m}{r}\right)+\frac{E^{2}}{c^2}\left(\frac{m \left(128 m^2-12 r^2 \left(\sqrt{1-\frac{2 m}{r}}-3\right)+m r \left(26 \sqrt{1-\frac{2 m}{r}}-99\right)\right)}{16 r^3 (r-2 m)^2}\right)\right.\notag\\
	&\left.+\left(\frac{l^{2}}{r^{2}}+hc^{2}\right)\left(\frac{m \left(48 m^2-4 r^2 \left(\sqrt{1-\frac{2 m}{r}}+1\right)+m r \left(4 \sqrt{1-\frac{2 m}{r}}-55\right)\right)}{16 r^4 (r-2 m)}\right)\right\}+\mathcal{O}(\Theta^{4}) \label{eq:4.7}
\end{align}

In the commutative limit ($\Theta \to 0$), the effective potential reduces to the standard Schwarzschild form \cite{Chandr1}.


\subsection{Radial motion of massive particles}\label{subsec:NCRM}

Here, we analyze the radial motion of a massive test particle in the NC Schwarzschild spacetime described by Eq.~\eqref{eq:4.1}. In this context, we set $\dot{\phi} = 0$ and $l = 0$, while assuming $h = 1$ and $c = 1$. Under these conditions, Eq.~\eqref{eqt 3.68} reduces to:
\begin{align}
	\left(\frac{dr}{d\tau} \right)^{2} &= E^2 - \left(1-\frac{2m}{r}\right) - \Theta^{2}\left\{E^{2}\left(\frac{m \left(128 m^2-12 r^2 \left(\sqrt{1-\frac{2 m}{r}}-3\right)+m r \left(26 \sqrt{1-\frac{2 m}{r}}-99\right)\right)}{16 r^3 (r-2 m)^2}\right)\right.\notag\\
	&\left.+\left(\frac{m \left(48 m^2-4 r^2 \left(\sqrt{1-\frac{2 m}{r}}+1\right)+m r \left(4 \sqrt{1-\frac{2 m}{r}}-55\right)\right)}{16 r^4 (r-2 m)}\right)\right\}+\mathcal{O}(\Theta^{4}) . 
	\label{eq:4.7'}
\end{align}
When $\Theta = 0$, this equation reduces to the classical radial motion equation of a massive particle in Schwarzschild spacetime.

The effective potential in case of radial time-like geodesics is given by:
\begin{align}
	V_{\text{eff}}(r,\Theta)&=\left(1-\frac{2 m}{r}\right)-E^{2}+\Theta^{2}\left\{E^{2}\left(\frac{m \left(128 m^2-12 r^2 \left(\sqrt{1-\frac{2 m}{r}}-3\right)+m r \left(26 \sqrt{1-\frac{2 m}{r}}-99\right)\right)}{16 r^3 (r-2 m)^2}\right)\right.\notag\\
	&\left.+\left(\frac{m \left(48 m^2-4 r^2 \left(\sqrt{1-\frac{2 m}{r}}+1\right)+m r \left(4 \sqrt{1-\frac{2 m}{r}}-55\right)\right)}{16 r^4 (r-2 m)}\right)\right\}+\mathcal{O}(\Theta^{4}) \label{eq:radialVeff}
\end{align}
\begin{figure}[h]
	\begin{center}
		\includegraphics[width=0.45\textwidth]{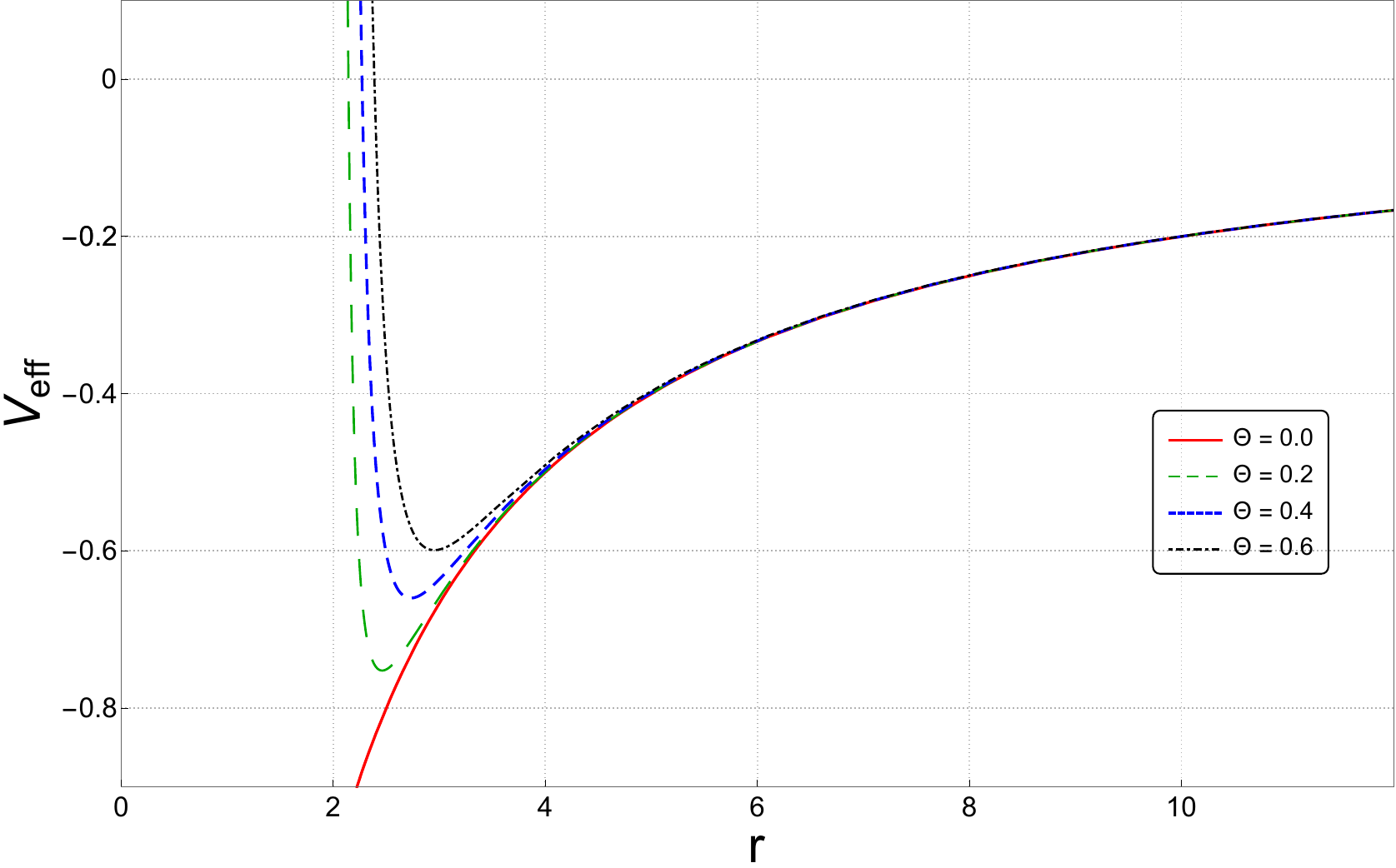}
		\includegraphics[width=0.45\textwidth]{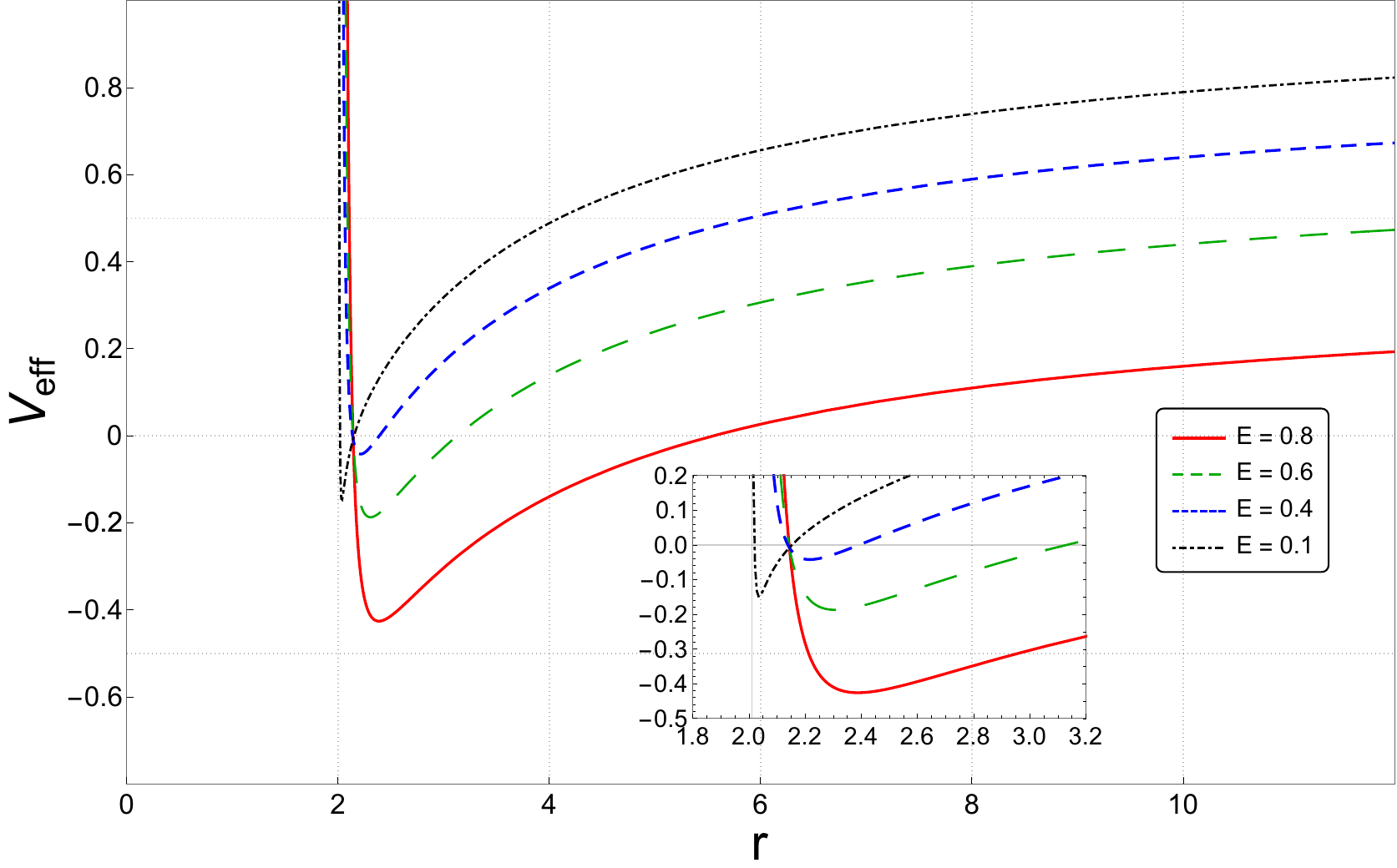}
	\end{center}
	\caption{Effective potential for a radial massive test particle as a function of $r$, for different values of $\Theta$ with fixed energy: $E = 1$, $m = 1$ (left panel), and for different energy $E$ with fixed $\Theta=0.2$ (right panel).}
	\label{fig:radial1}
\end{figure}
In Fig.~\ref{fig:radial1}, we illustrate the behavior of the effective potential for the radial trajectory ($l=0$) of a massive test particle in NC spacetime as a function of $r$. The left and right panels respectively depict variations for different values of the NC parameter and different energy levels of the particle. It is evident that, even in radial motion, non-commutativity prevents a massive test particle with high energy ($E=1$) from reaching the NC event horizon by predicting a new minimum outside the NC event horizon. Moreover, when we fix the NC parameter and vary the energy of the particle, an interesting behavior emerges: a particle with low energy can fall into the NC event horizon but cannot reach the NC singularity, which in this geometry is shifted to a finite radius, $r_{\text{singularity}}^{NC} = 2m$. For a very low energy ($E \ll 1$), this phenomenon disappears, and the massive test particle falls into the NC singularity at the finite radius $2m$ (see Ref.~\cite{abdellah1}).

Next, we consider the free-fall motion of a massive particle in the NC Schwarzschild BH geometry. The particle is initially at rest, i.e., $\dot{r} = 0$ at $r = r_0$ when $\tau = 0$. For the proper time in NC spacetime, we set $E = 1$ in Eq.~\eqref{eq:4.7'} and integrate, yielding:
\begin{align}
	\hat{\tau} &= -\int_{r_0}^{r} \left(\frac{2m}{r'} - \Theta^{2} \left\{\left(\frac{m \left(128 m^2-12 r^2 \left(\sqrt{1-\frac{2 m}{r}}-3\right)+m r \left(26 \sqrt{1-\frac{2 m}{r}}-99\right)\right)}{16 r^3 (r-2 m)^2}\right)\right.\right.\notag\\
	&\quad \left.\left. + \left(\frac{m \left(48 m^2-4 r^2 \left(\sqrt{1-\frac{2 m}{r}}+1\right)+m r \left(4 \sqrt{1-\frac{2 m}{r}}-55\right)\right)}{16 r^4 (r-2 m)}\right) \right\}\right)^{-1/2} dr'.\notag\\
	&= -\int_{r_0}^{r} \sqrt{\frac{2m}{r'}} \left(1 - \Theta^{2}\frac{r'}{2m} \left\{\left(\frac{m \left(128 m^2-12 r^2 \left(\sqrt{1-\frac{2 m}{r}}-3\right)+m r \left(26 \sqrt{1-\frac{2 m}{r}}-99\right)\right)}{16 r^3 (r-2 m)^2}\right)\right.\right.\notag\\
	&\quad \left.\left. + \left(\frac{m \left(48 m^2-4 r^2 \left(\sqrt{1-\frac{2 m}{r}}+1\right)+m r \left(4 \sqrt{1-\frac{2 m}{r}}-55\right)\right)}{16 r^4 (r-2 m)}\right)\right\}\right)^{-1/2} dr'.
	\label{eq:4.71'}
\end{align}
we choose:
\begin{align}
	\epsilon &= \Theta^{2}\frac{r'}{2m} \left\{\left(\frac{m \left(128 m^2-12 r^2 \left(\sqrt{1-\frac{2 m}{r}}-3\right)+m r \left(26 \sqrt{1-\frac{2 m}{r}}-99\right)\right)}{16 r^3 (r-2 m)^2}\right)\right.\notag\\
	&\quad \left. + \left(\frac{m \left(48 m^2-4 r^2 \left(\sqrt{1-\frac{2 m}{r}}+1\right)+m r \left(4 \sqrt{1-\frac{2 m}{r}}-55\right)\right)}{16 r^4 (r-2 m)}\right)\right\},
	\label{eq:25a}
\end{align}
where $\epsilon$ is a small parameter due to the smallness of $\Theta$. 
Here, we use the first-order approximation of $\epsilon$:
\begin{equation}
	\left( 1-\epsilon \right) ^{-1/2} \approx 1 + \frac{1}{2} \epsilon + \mathcal{O}\left(\epsilon^2\right),
	\label{eq:25b}
\end{equation}

Using  Eq.~\eqref{eq:25b}, proper time in equation \eqref{eq:4.71'} can be written up to the second order of $\Theta$ as
\begin{align}
	\hat{\tau} &= -\int_{r_0}^{r} \left(\sqrt{\frac{r'}{2m}} - \Theta^{2} \sqrt{\frac{r'}{2m}} \left\{\frac{48m^3-143m^2r'+73mr'^2-16r'^3+(4m^2r'-19mr'^2+8r'^3)\sqrt{1-\frac{2m}{r'}}}{32r'^5(1-\frac{2m}{r'})^2}\right\} \right) dr'.
	\label{eq:4.71b'}
\end{align}

After integration, the proper time expression up to the second order in $\Theta$ is:
\begin{align}
	\hat{\tau} &= \frac{2}{3}\left(\sqrt{\frac{r_0^3}{2m}} - \sqrt{\frac{r^3}{2m}}\right) + \Theta^2 \left\{ \sqrt{\frac{r_0}{2m}} \left(\frac{64m^2+4m(-103+2\sqrt{1-\frac{2m}{r_0}})r_0+227r_0^2}{128(2m-r_0)r_0^2}\right)\notag\right.\\
	&\quad \left. - \sqrt{\frac{r}{2m}} \left(\frac{64m^2+4m(-103+2\sqrt{1-\frac{2m}{r}})r+227r^2}{128(2m-r)r^2}\right) + \frac{1}{4m} \left(\text{ArcSin}\left(\sqrt{\frac{2m}{r_0}}\right) - \text{ArcSin}\left(\sqrt{\frac{2m}{r}}\right)\right)\right.\notag\\
	&\quad \left. + \frac{99}{256m} \left(\text{ArcTanh}\left(\sqrt{\frac{r_0}{2m}}\right) - \text{ArcTanh}\left(\sqrt{\frac{r}{2m}}\right)\right) \right\}.
	\label{eq:4.72'}
\end{align}

In the limit $\Theta = 0$, this expression reduced to the commutative case, recovering the classical result.

\begin{figure}[h]
	\begin{center}
		\includegraphics[width=0.5\textwidth]{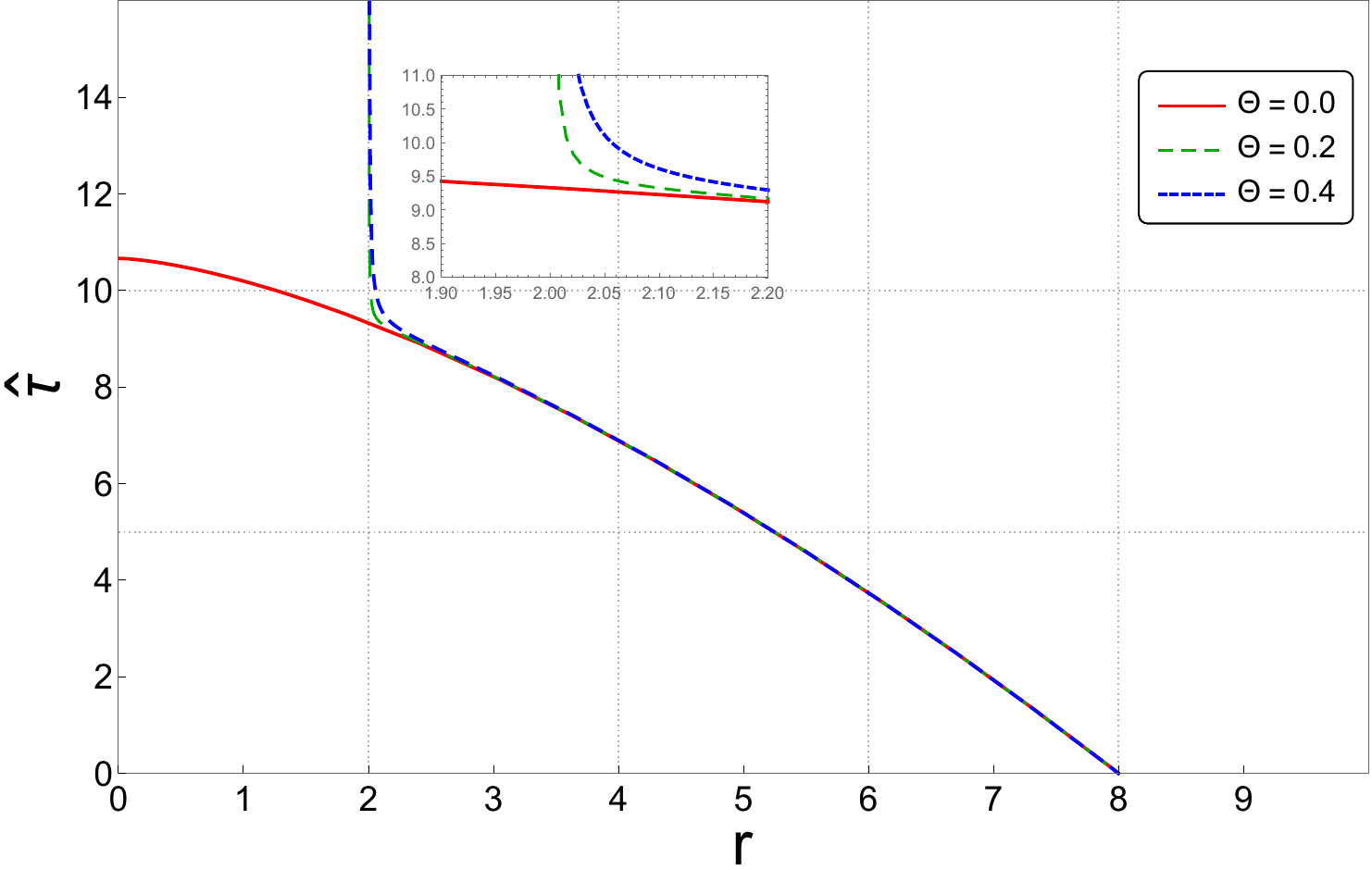}
	\end{center}
	\caption{Proper time evolution for the free fall of a massive particle in the NC Schwarzschild BH as a function of $r$, with $m=1$ and the initial position $r_0=8$.}\label{figa}
\end{figure}

Fig. \ref{figa} illustrates the variation of proper time $\hat{\tau}$ as a function of $r$ during the free fall of a massive test particle into the NC Schwarzschild BH. The NC parameter $\Theta$ significantly influences the proper time near the event horizon, increasing its duration. An infinite amount of time is required to reach the NC singularity due to non-commutativity effects (see Ref.~\cite{abdellah1}), a phenomenon that is absent in the commutative case \cite{Chandr1}. 

Similarly to proper time, the coordinate time $\hat{t}$ for free fall in the NC Schwarzschild BH can be obtained using Eqs.~\eqref{eq:4.5} and \eqref{eq:4.7'} together with $E = 1$:

\begin{align}
	\hat{t}&=-\int_{r_0}^{r}\left(\hat{g}_{tt}\right)^{-1}\left(\frac{2 m}{r'}-\Theta^{2}\left\{\left(\frac{m \left(128 m^2-12 r^2 \left(\sqrt{1-\frac{2 m}{r}}-3\right)+m r \left(26 \sqrt{1-\frac{2 m}{r}}-99\right)\right)}{16 r^3 (r-2 m)^2}\right)\right.\right.\notag\\
	&\quad \left.\left. + \left(\frac{m \left(48 m^2-4 r^2 \left(\sqrt{1-\frac{2 m}{r}}+1\right)+m r \left(4 \sqrt{1-\frac{2 m}{r}}-55\right)\right)}{16 r^4 (r-2 m)}\right) \right\}\right)^{-1/2}dr'\, , \label{eq:4.73'}
\end{align}

At leading order in $\Theta$, and integration, the coordinate time is given by:
\begin{align}
	\hat{t}&= \frac{2}{3}\left(\sqrt{\frac{r_0^3}{2m}}-\sqrt{\frac{r^3}{2m}}\right)+4m \left( \sqrt{\frac{r_0}{2m}}-\sqrt{\frac{r}{2m}}\right)-4 m \left(\text{ArcTanh}\left(\sqrt{\frac{r_0}{2m}}\right)-\text{ArcTanh}\left(\sqrt{\frac{r}{2m}}\right)\right)\notag\\
	&+\Theta^2\left\{\sqrt{\frac{r}{2m}}\left(-\frac{3211 m}{256(r-2m)^2}+\frac{25m^2}{2r(r-2m)^2}+\frac{1693r}{512(r-2m)^2}+\frac{(8r-15m) \sqrt{1-\frac{2m}{r}}}{16(r-2m)^2}\right)\notag\right.\\
	&\left.-\sqrt{\frac{r_0}{2m}}\left(-\frac{3211 m}{256(r_0-2m)^2}+\frac{25m^2}{2r_0(r_0-2m)^2}+\frac{1693r_0}{512(r_0-2m)^2}+\frac{(8r_0-15m) \sqrt{1-\frac{2m}{r_0}}}{16(r_0-2m)^2}\right)\right.\notag\\
	&\left.+\frac{1}{2m}\left(\text{ArcSin}\left(\sqrt{\frac{2m}{r_0}}\right)-\text{ArcSin}\left(\sqrt{\frac{2m}{r}}\right)\right)+\frac{1181}{1024m}\left(\text{ArcTanh}\left(\sqrt{\frac{r_0}{2m}}\right)-\text{ArcTanh}\left(\sqrt{\frac{r}{2m}}\right)\right)\right\}.
	\label{eq:4.74'}
\end{align}
The commutative expression for the coordinate time is recovered by setting $\Theta=0$.

\begin{figure}[h]
	\begin{center}
		\includegraphics[width=0.5\textwidth]{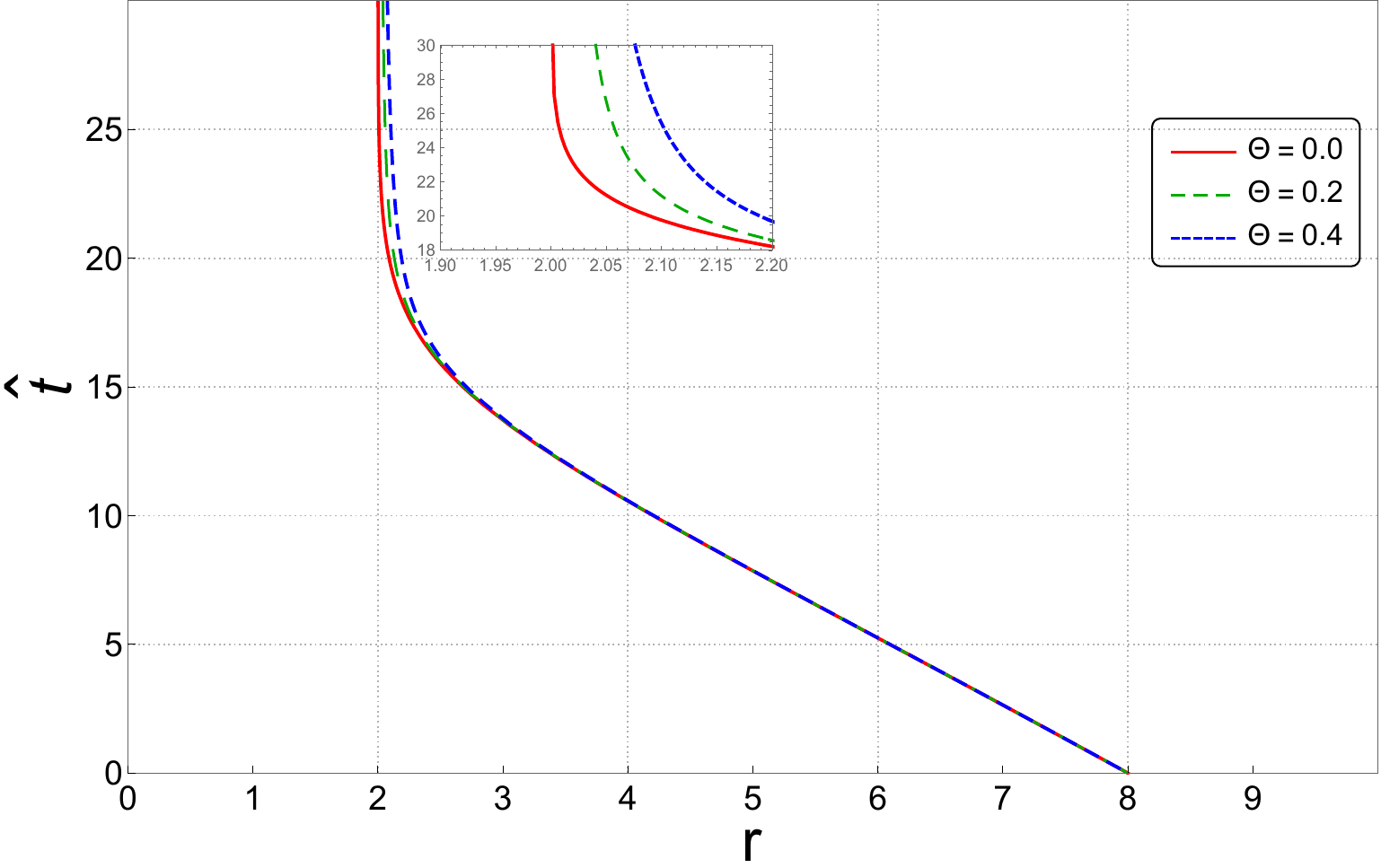}
	\end{center}
	\caption{Coordinate time evolution for the free fall of a massive particle in the NC Schwarzschild BH as a function of $r$, with $m=1$ and the initial position $r_0=8$.}\label{figb}
\end{figure}

In Fig.~\ref{figb}, we present the variation of coordinate time $\hat{t}$ during the free fall of a massive test particle in the NC Schwarzschild BH. The effect of non-commutativity becomes significant near the NC event horizon, where it leads to an increase in the coordinate time required for the particle to approach the NC horizon. This qualitative behavior aligns with the commutative case, as discussed in \cite{Chandr1}, but is further modified due to the presence of non-commutativity.

\begin{figure}[h]
	\begin{center}
		\includegraphics[width=0.5\textwidth]{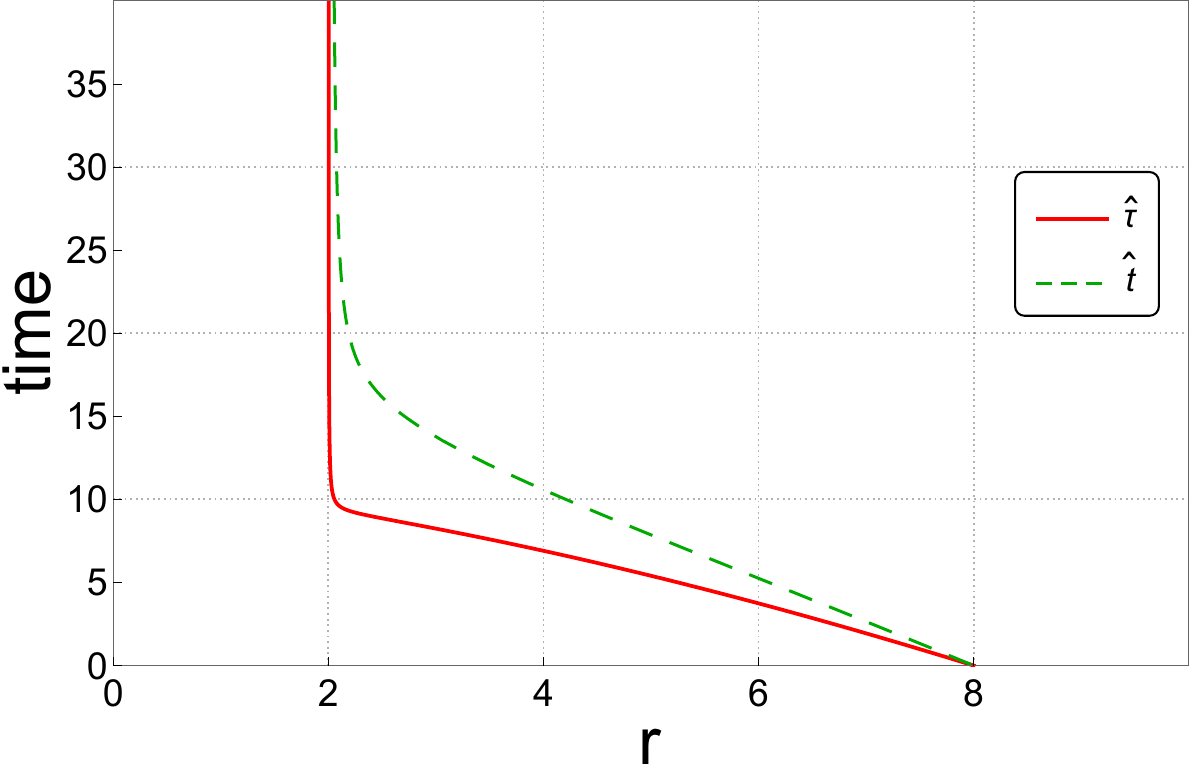}
	\end{center}
	\caption{Comparison of proper and coordinate time for the free fall of a massive particle in the NC Schwarzschild BH as a function of $r$, with $m=1$ and the initial position $r_0=8$.}\label{figc}
\end{figure}

Fig.~\ref{figc} depicts the behavior of both proper time and coordinate time for a massive particle falling toward the NC Schwarzschild BH. These results indicate that non-commutativity prevents a massive particle from reaching the NC singularity within a finite time, in contrast to the commutative case \cite{Chandr1}, thus confirming our analysis of the effective potential of the radial trajectory of a massive test particle. It is noteworthy that in this approach, the commutative case is recovered in the limit $\Theta=0$, preserving the expected large-distance commutative behavior while introducing novel effects near the event horizon. This feature is not addressed in other approaches related to the NC matter distribution framework. In some cases, these approaches lose all commutative behavior (e.g., Refs. \cite{rahaman,bhar}), while in others, they preserve the general commutative behavior but predict a reduction in proper time \cite{larranaga}. This contrasts with our results, where this geometry ensures that the particle requires infinite proper time to reach the NC singularity. Furthermore, the results highlight the impact of utilizing the gauge theory of gravity in NC spacetime to analyze the geodesic motion of test particles.


\subsection{Circular motion of massive particles }\label{subsec:NCCM}

The effective potential governing the dynamics of a massive test particle is given by \eqref{eq:4.7} for $h=1$, $l \neq 0$, and $c=1$, and is analyzed in detail in Ref. \cite{abdellah1}.
\begin{figure}[h]
	\begin{center}
		\includegraphics[width=0.5\textwidth]{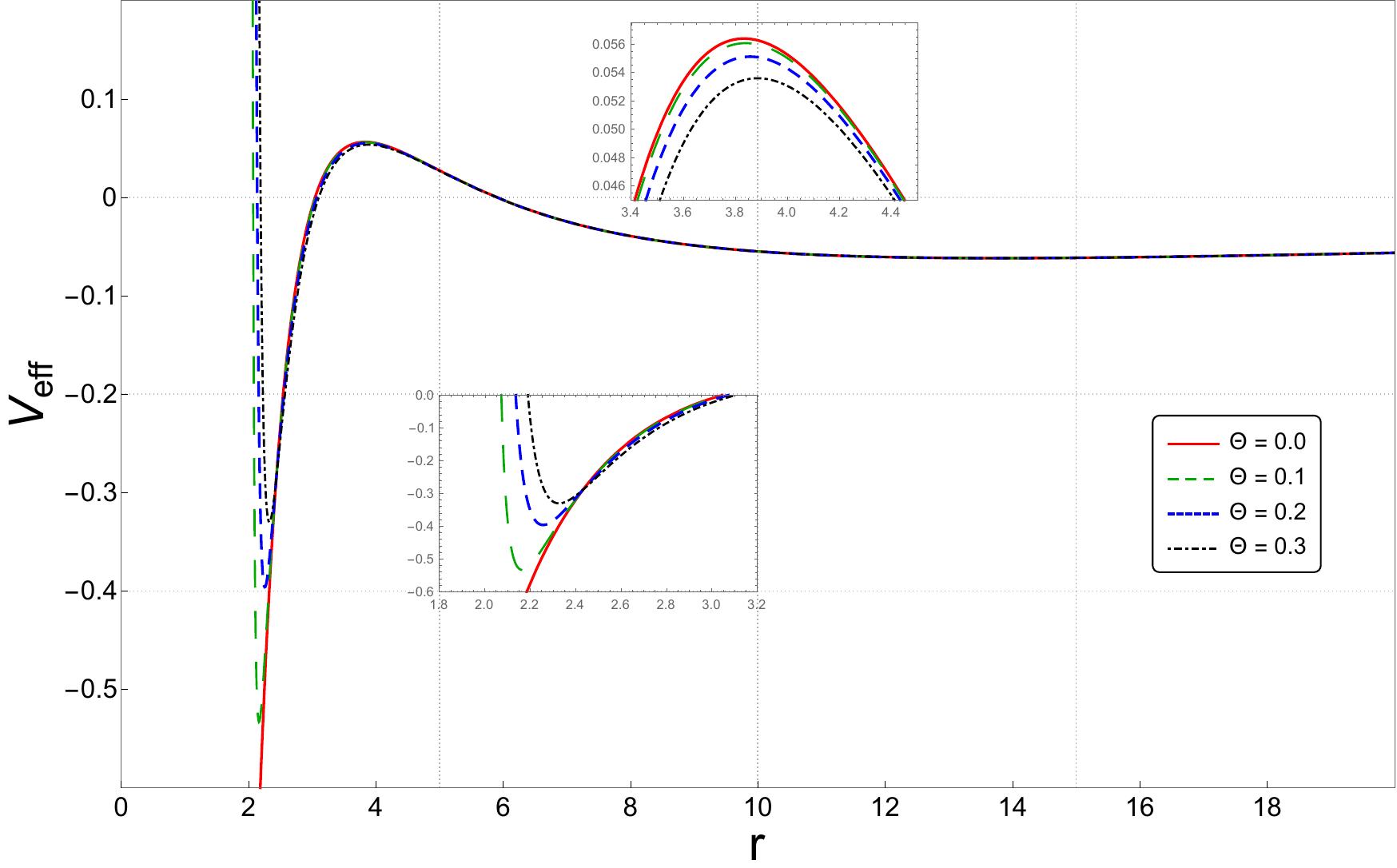}
	\end{center}
	\caption{Effective potential for a massive particle as a function of $r$, for different values of $\Theta$ with fixed parameters: $E = 0.998$, $m = 1$, and $l = 4.2$.}
	\label{fig1}
\end{figure}

Fig.~\ref{fig1} illustrates the influence of the NC parameter $\Theta$ on the effective potential of a massive test particle. Notably, the introduction of non-commutativity generates a new minimum in the effective potential near the event horizon, regardless of the angular momentum values ($l \geq 0$). This new minimum, located outside the event horizon, corresponds to a stable circular orbit \cite{abdellah1}, leading to the existence of multiple stable circular orbits in NC Schwarzschild geometry. 

In contrast, prior studies on non-commutativity in geodesic motion whether based on NC matter distributions \cite{nozari1,nozari2,kuniyal,iftikhar,larranaga,bhar,rahaman,NCGD1,NCGD3,NCLD2,NCLD3}, NC gauge theory of gravity via deformed geometry \cite{NCGTG2,NCGTG3}, or deformed mass \cite{NCGTGmassdeformed1,NCGTGmassdeformed2,NCGTGmassdeformed3} have not predicted these new stable circular orbits near the event horizon.

To assess the stability of these circular orbits, the effective potential must satisfy the following conditions:
\begin{equation}\label{eq:27}
	V_{eff}(r, \Theta) = 0, \quad \frac{d V_{eff}}{d r} = 0.
\end{equation}

Using these conditions, the leading-order corrections in $\Theta$ for the energy $E_c^2$ and angular momentum $L_c^2$ of circular orbits are given by:

\begin{align}
	E^2_c &\simeq \frac{(r_c-2m )^2}{r_c (r_c - 3m)} - \frac{\left(B(r_c) + F(r_c)\sqrt{1 - \frac{2m}{r_c}}\right)\Theta^2}{8 r_c^4 (-3m + r_c)^2 \sqrt{1 - \frac{2m}{r_c}}} + \mathcal{O}(\Theta^4), \label{eq:4.9} \\
	L^2_c &\simeq \frac{mr_c^2}{r_c - 3m} - \frac{\left(S(r_c) + Q(r_c)\sqrt{1 - \frac{2m}{r_c}}\right)\Theta^2}{8 r_c (r_c-3m )^2 (r_c - 2m) \sqrt{1 - \frac{2m}{r_c}}} + \mathcal{O}(\Theta^4), \label{eq:4.10}
\end{align}
where the coefficients are defined as:
\begin{subequations}
	\begin{align}
		B(r_c) &= -50m^4 + 58m^3r_c - 18m^2r_c^2 - mr_c^3, \quad F(r_c) = -96m^4 - 61m^3r_c - m^2r_c^2 + 3mr_c^3, \\
		S(r_c) &= -2m^4 + 6m^3r_c +8m^2r_c^2-6mr_c^3, \quad Q(r_c) = -72m^4 + 148m^3r_c - 85m^2r_c^2 + 14mr_c^3.
	\end{align}
\end{subequations}
In the limit $\Theta \to 0$, these expressions reduce to their commutative counterparts for Schwarzschild spacetime.

\begin{figure}[h]
	\begin{center}
		\includegraphics[width=0.48\textwidth]{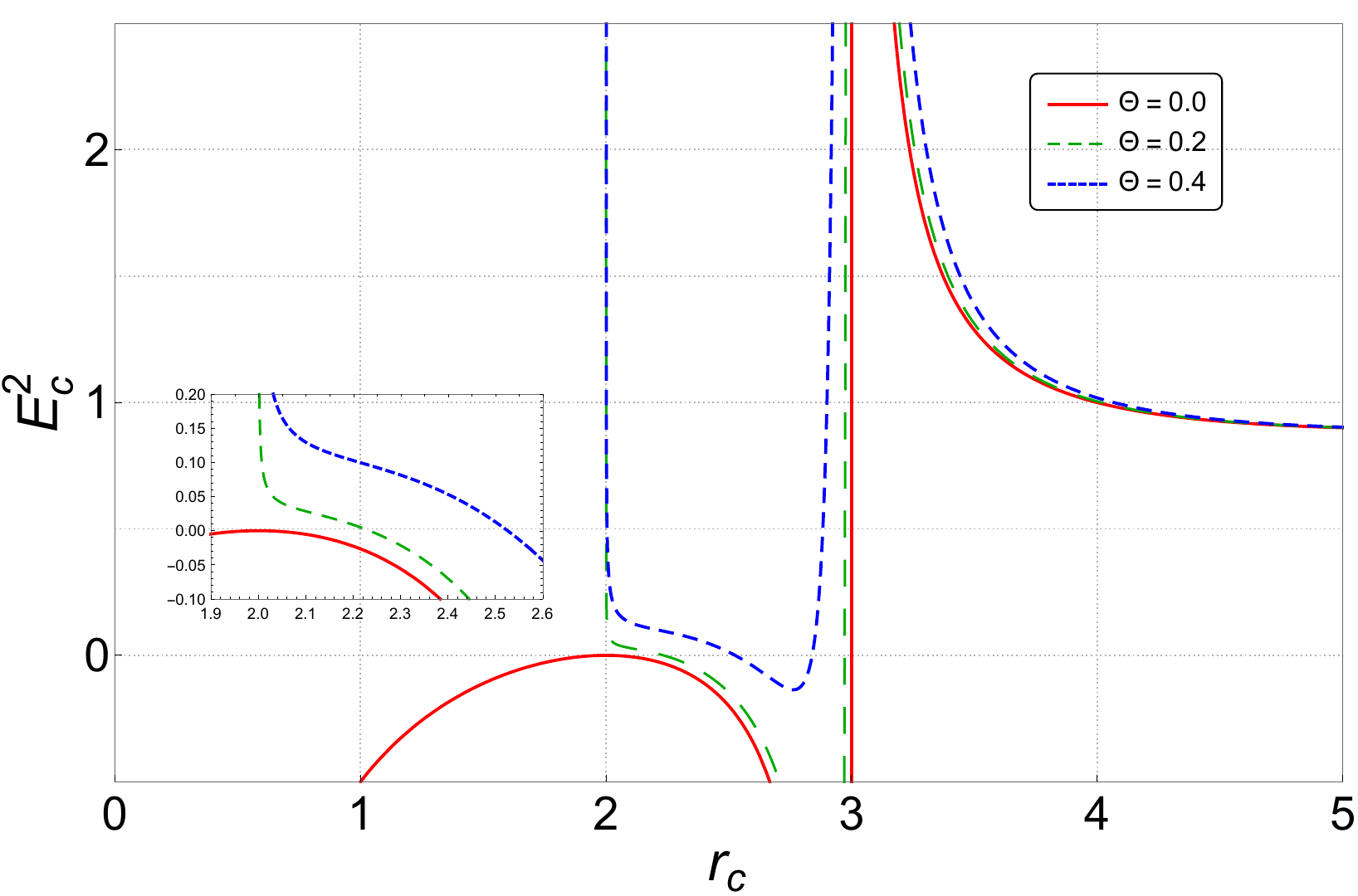}\hfill
		\includegraphics[width=0.5\textwidth]{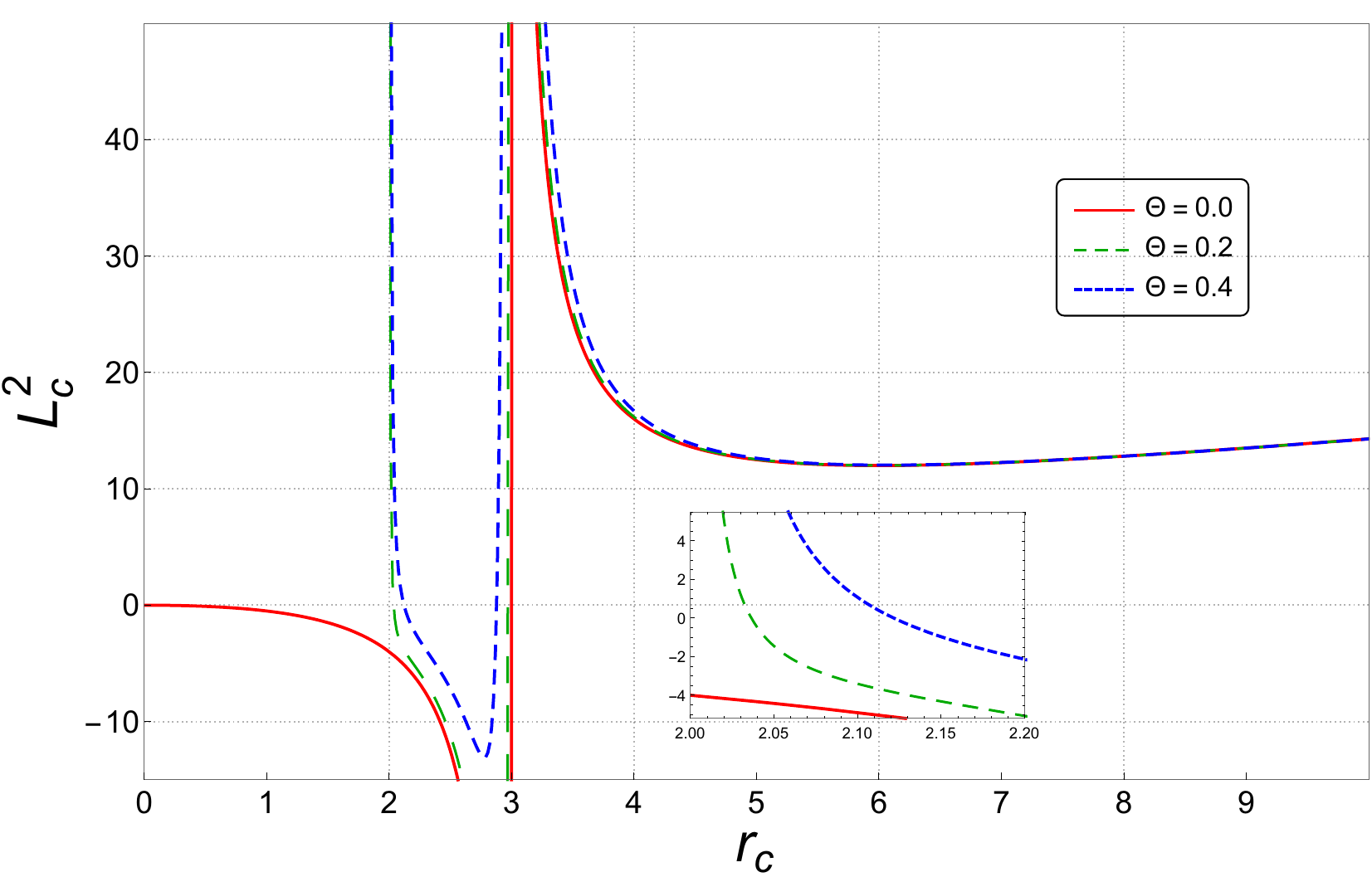}
	\end{center}
	\caption{Energy $E_c^2$ (left panel) and angular momentum $L_c^2$ (right panel) for circular orbits of massive particles in NC Schwarzschild spacetime, as functions of $r_c$.}
	\label{fig2}
\end{figure}

Fig.~\ref{fig2} presents the behavior of $E_c^2$ and $L_c^2$ for circular orbits as functions of $r_c$, highlighting the impact of the NC parameter $\Theta$. The left panel shows that NC modifications introduce a region near the event horizon where the energy condition becomes unphysical ($E_c^2 < 0$). However, beyond this region, $E_c^2$ transitions into a physical region ($E_c^2 > 0$), indicating the emergence of a new innermost stable circular orbit (ISCO) in NC geometry \cite{abdellah1}. In the right panel, the influence of $\Theta$ on angular momentum is evident, with positive values corresponding to the newly formed stable circular orbits near the event horizon. When $\Theta = 0$, these additional features disappear, restoring the classical Schwarzschild behavior.

\subsubsection{Orbital velocity and time period}

The orbital velocity of a massive test particle in the NC Schwarzschild spacetime is defined as \cite{geostab3,geostab5}:
\begin{equation}\label{eq:omega}
	\hat{\Omega}_c = \frac{\dot{\phi}}{\dot{t}}.
\end{equation}
Utilizing Eqs.~\eqref{eq:4.5}, \eqref{eq:4.9}, and \eqref{eq:4.10}, the orbital velocity, accurate to second order in the NC parameter $\Theta$, is given by:
\begin{equation}
	\hat{\Omega}_c = \sqrt{\frac{m}{r_c^3} + \frac{X(r_c) + W(r_c)\sqrt{1 - \frac{2m}{r_c}}}{8 r_c^6 (r_c - 2m)^3}\Theta^2},
\end{equation}
where the functions $X(r_c)$ and $W(r_c)$ are defined as:
\begin{align}
	X(r_c) &= -640 m^5 + 1158 m^4r_c - 775 m^3r_c^2 + 226 m^2r_c^3 - 24 mr_c^4, \notag\\
	W(r_c) &= -98 m^4r_c + 148 m^3r_c^2 - 69 m^2r_c^3 + 12 mr_c^4.
\end{align}
In the limit $\Theta = 0$, the commutative Schwarzschild case is recovered \cite{geostab3}.

The time period, or orbital times scale of the coordinate time in NC spacetime is given by \cite{geostab3}:
\begin{equation}\label{eq:period}
	\hat{T}_{\hat{\Omega}} = \frac{2\pi}{\hat{\Omega}_c}.
\end{equation}

\begin{figure}[h]
	\begin{center}
		\includegraphics[width=0.48\textwidth]{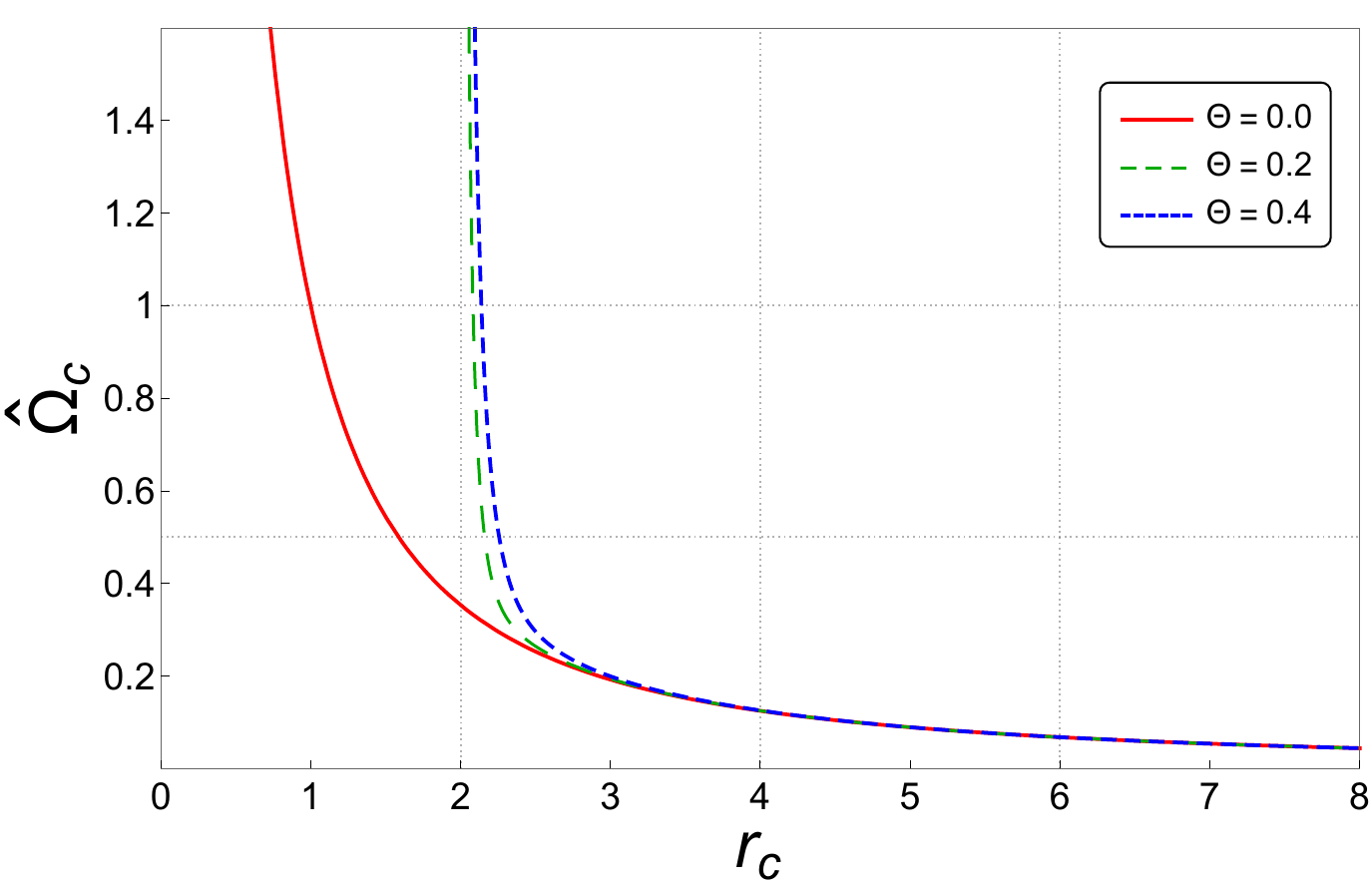}\hfill
		\includegraphics[width=0.48\textwidth]{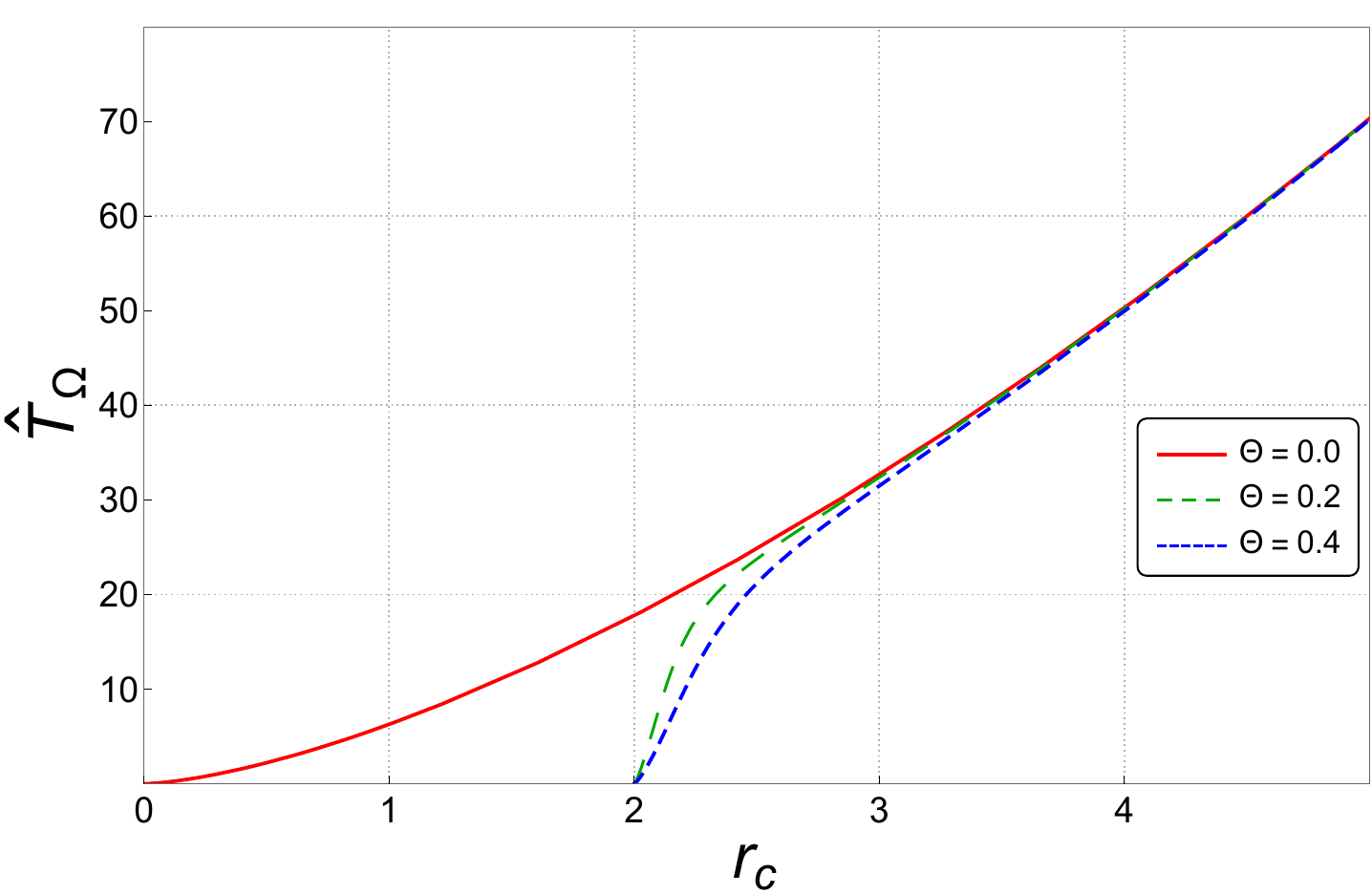}
	\end{center}
	\caption{The behavior of the NC orbital velocity (left panel) and the NC orbital period (right panel) for a massive test particle.}
	\label{fig3}
\end{figure}

Fig.~\ref{fig3} illustrates the orbital velocity (left panel) and the orbital time scale (right panel) of a massive test particle in NC Schwarzschild spacetime as functions of $r_c$. The effects of non-commutativity become significant near the NC event horizon $r_h^{NC}$. Specifically, non-commutativity increases the orbital velocity in the vicinity of $r_h^{NC}$, while its impact becomes negligible at larger orbital radii. Similarly, the orbital period decreases due to non-commutativity, approaching zero near the NC event horizon. This reduction is most pronounced close to $r_h^{NC}$ and becomes negligible for $r_c \gg 3$, where NC effects vanish.

\subsubsection{Lyapunov exponents}

To analyze the stability and instability of circular geodesics for a massive test particle in the NC Schwarzschild spacetime, we employ the Lyapunov exponents $\hat{\lambda}$. The stability of an orbit is determined by the nature of $\hat{\lambda}$: an imaginary $\hat{\lambda}$ indicates a stable orbit, whereas a real $\hat{\lambda}$ signifies instability \cite{geostab2,geostab10}. The proper-time ($\hat{\lambda}_p$) and coordinate-time ($\hat{\lambda}_c$) Lyapunov exponents in NC spacetime are given by \cite{geostab3,geostab2,geostab10,geostab1}:
\begin{equation}
	\hat{\lambda}_p = \pm\sqrt{\frac{-V_{eff}''(r, \Theta)}{2}}, \quad 
	\hat{\lambda}_c = \pm\sqrt{\frac{-V_{eff}''(r, \Theta)}{2\dot{t}^2}}.
\end{equation}

Using the effective potential \eqref{eq:4.7} along with Eqs. \eqref{eq:4.5}, \eqref{eq:4.9}, and \eqref{eq:4.10}, the Lyapunov exponents are expressed as:
\begin{align}
	\hat{\lambda}_p &= \sqrt{-\frac{m(r_c - 6m)}{(r_c - 3m) r_c^3} + 
		\frac{Z(r_c) + P(r_c)\sqrt{1 - \frac{2m}{r_c}}}{16(r_c - 2m)^3 r_c^6 (r_c - 3m)^2}\Theta^2}, \label{eq:lamp} \\
	\hat{\lambda}_c &= \sqrt{-\frac{m(r_c - 6m)}{r_c^4} + 
		\frac{Y(r_c) + N(r_c)\sqrt{1 - \frac{2m}{r_c}}}{16(r_c - 2m)^3 r_c^7}\Theta^2}. \label{eq:lamc}
\end{align}

The functions $Z(r_c)$, $P(r_c)$, $Y(r_c)$, and $N(r_c)$ are given by:
\begin{subequations}
	\begin{align}
		Z(r_c) &= 15552 m^7 - 34188 m^6r_c + 28548 m^5r_c^2 - 11285 m^4r_c^3 + 2043 m^3r_c^4 - 102 m^2r_c^5 - 8 mr_c^6, \\
		P(r_c) &= 2904 m^6r_c - 6196 m^5r_c^2 + 5041 m^4r_c^3 - 1851 m^3r_c^4 + 256 m^2r_c^5, \\
		Y(r_c) &= 8640 m^6 - 15436 m^5r_c + 9744 m^4r_c^2 - 2517 m^3r_c^3 + 196 m^2r_c^4 + 8 mr_c^5, \\
		N(r_c) &= 1488 m^5r_c - 2408 m^4r_c^2 + 1367 m^3r_c^3 - 290 m^2r_c^4.
	\end{align}
\end{subequations}

In the limit $\Theta = 0$, the commutative Schwarzschild case is recovered \cite{geostab1,geostab10}.

\begin{figure}[h]
	\begin{center}
		\includegraphics[width=0.48\textwidth]{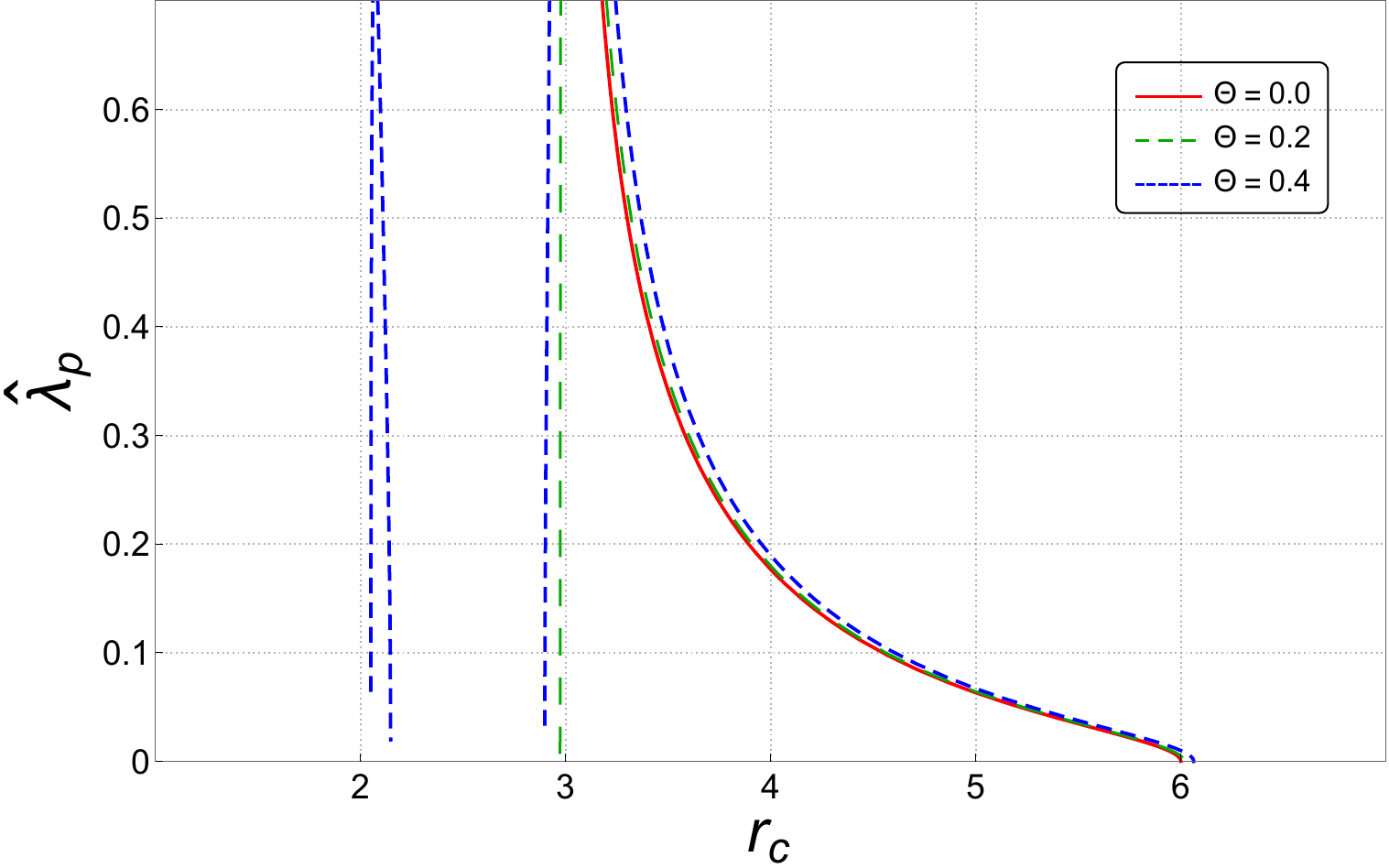}\hfill
		\includegraphics[width=0.48\textwidth]{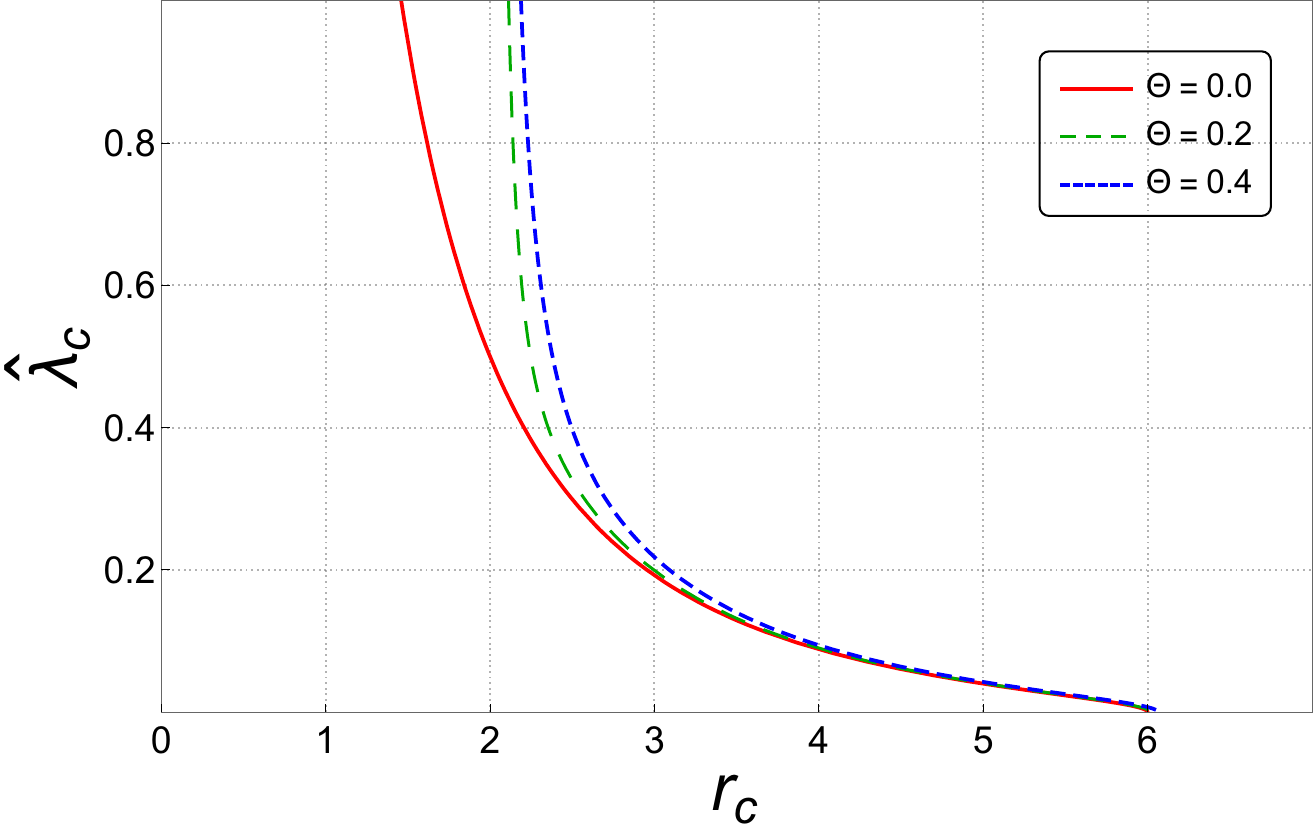}
	\end{center}
	\caption{Proper time (left panel) and coordinate time (right panel) Lyapunov exponents for a massive particle as functions of the circular orbit radius $r_c$.}
	\label{fig4}
\end{figure}

Fig. \ref{fig4} illustrates the proper-time (left panel) and coordinate-time (right panel) Lyapunov exponents for a massive test particle in NC spacetime. The coordinate-time Lyapunov exponent $\hat{\lambda}_c$ is real in the region $r_h^{NC} < r_c \leq r_{outer}^{ISCO}$, indicating unstable orbits near the event horizon. As the orbital radius $r_c$ increases, the instability decreases, reaching zero at the outer ISCO, $r_{outer}^{ISCO}$. The proper-time Lyapunov exponent $\hat{\lambda}_p$ exhibits distinct behavior: instability (real values) appears in two regions, $r_h^{NC} < r_c \leq r_{inner}^{ISCO}$ and $r_c^{uns} < r_c \leq r_{outer}^{ISCO}$. However, in the intermediate range $r_{inner}^{ISCO} \leq r_c \leq r_c^{uns}$, $\hat{\lambda}_p$ becomes imaginary, indicating stable circular orbits. These findings confirm the existence of new stable circular orbits near the event horizon, consistent with previous results \cite{abdellah1}.

The ratio of the proper-time to coordinate-time Lyapunov exponents is given by:

\begin{equation}
	\frac{\hat{\lambda}_p}{\hat{\lambda}_c} = 
	\sqrt{\frac{r_c}{r_c - 3m} + 
		\frac{Y(r_c)(3m - r_c) - Z(r_c) + 
			(N(r_c)(3m - r_c) - P(r_c))\sqrt{1 - \frac{2m}{r_c}}}{16m(r_c - 2m)^3(r_c - 3m)^2(r_c - 6m)r_c^2}\Theta^2}.
\end{equation}
In the limit $\Theta = 0$, the commutative case is recovered \cite{geostab1,geostab10}.

\begin{figure}[h]
	\begin{center}
		\includegraphics[width=0.5\textwidth]{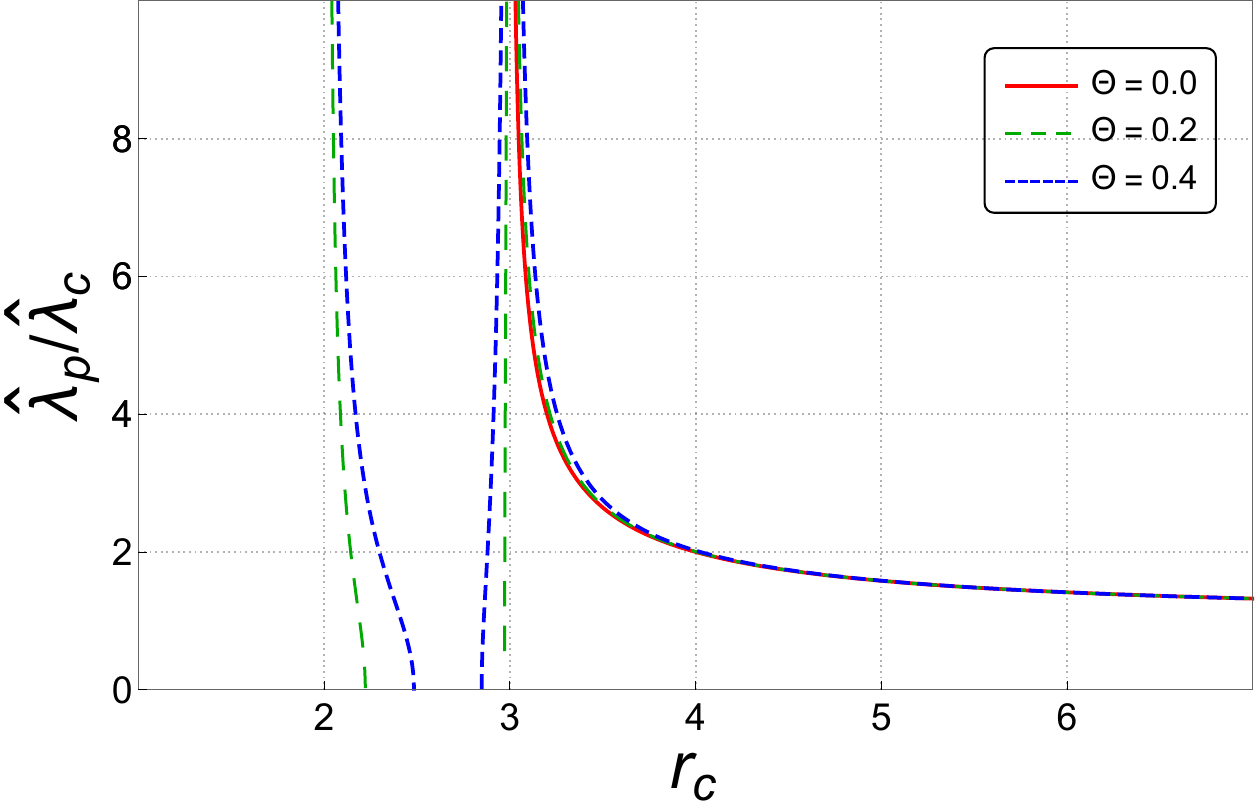}
	\end{center}
	\caption{The ratio of proper time to coordinate time Lyapunov exponents for a massive test particle in NC Schwarzschild spacetime as a function of $r_c$.}
	\label{fig5}
\end{figure}

Fig. \ref{fig5} illustrates the ratio $\hat{\lambda}_p / \hat{\lambda}_c$ as a function of $r_c$ in NC spacetime. This ratio remains real in two distinct regions, separated by an interval where it becomes imaginary, mirroring the behavior observed in the proper-time Lyapunov exponent. The analysis of proper-time and coordinate-time Lyapunov exponents, along with their ratio, reveals a novel feature in the dynamics of circular motion around an NC Schwarzschild BH: the existence of a new range of stable circular orbits interspersed between unstable ones.

\subsection{Radial motion of massless particles}

In this subsection, we study the free-fall motion of a massless test particle in this geometry. The motion is described by the radial trajectory of a massless particle in the NC Schwarzschild spacetime and is obtained using the effective potential \eqref{eq:4.7}, setting $h = 0$, $c = 0$, and $l = 0$. This leads to the equation governing radial motion:
\begin{align}
	\left(\frac{dr}{d\tau} \right)^{2} &= E^2 - \Theta^{2} \left\{ E^{2}\left(\frac{m \left(128 m^2-12 r^2 \left(\sqrt{1-\frac{2 m}{r}}-3\right)+m r \left(26 \sqrt{1-\frac{2 m}{r}}-99\right)\right)}{16 r^3 (r-2 m)^2}\right) \right\} + \mathcal{O}(\Theta^4), \label{eq:4.8'}
\end{align}
where the commutative case is recovered for $\Theta = 0$. For a photon freely falling toward the NC Schwarzschild BH, emitted from a radial position $r = r_0$ with energy $E = 1$, the affine parameter $\tau$ and coordinate time $\hat{t}$ are given by:
\begin{align}
	\hat{\tau} &= -\int_{r_0}^{r} \left[ 1 - \Theta^{2} \left(\frac{m \left(128 m^2-12 r^2 \left(\sqrt{1-\frac{2 m}{r}}-3\right)+m r \left(26 \sqrt{1-\frac{2 m}{r}}-99\right)\right)}{16 r^3 (r-2 m)^2}\right) \right]^{-\frac{1}{2}} dr, \label{eq:4.81'}\\
	\hat{t} &= -\int_{r_0}^{r} (\hat{g}_{00})^{-1} \left[ 1 - \Theta^{2} \left(\frac{m \left(128 m^2-12 r^2 \left(\sqrt{1-\frac{2 m}{r}}-3\right)+m r \left(26 \sqrt{1-\frac{2 m}{r}}-99\right)\right)}{16 r^3 (r-2 m)^2}\right) \right]^{-\frac{1}{2}} dr. \label{eq:4.82'}
\end{align}

By evaluating these integrals to leading order in $\Theta$, the results are:
\begin{align}
	\hat{\tau} &= r_0 - r + \frac{\Theta^2}{128m} \bigg\{ 33\ln\left(\frac{r_0(r-2m)}{r(r_0-2m)}\right) 
	- \left[ \frac{m (64 m+29 r_0)+16 r_0^2}{r_0^2} - \frac{m (64 m+29 r)+16 r^2}{r^2} \right] \notag \\
	&\quad + \bigg[ 2\Bigg(\left(1 - \frac{2m}{r}\right)^{-\frac{1}{2}} - \left(1 - \frac{2m}{r_0}\right)^{-\frac{1}{2}} \Bigg)
	+ 28\bigg(\sqrt{1-\frac{2m}{r}} - \sqrt{1-\frac{2m}{r_0}}\bigg) \notag \\
	&\quad + \frac{26}{3} \bigg(\bigg(1-\frac{2m}{r_0}\bigg)^{\frac{3}{2}} - \bigg(1-\frac{2m}{r}\bigg)^{\frac{3}{2}}\bigg) 
	- \frac{37}{2} \frac{r_0(r-2m) - r(r_0-2m)}{(r_0-2m)(r-2m)} \bigg] \bigg\}, \label{eq:4.83'}\\
	\hat{t} &= r_0 - r + 2m\ln\left(\frac{r_0-2m}{r-2m}\right) + \frac{\Theta^2}{128m} \bigg\{ \frac{r^2 (104 m-11 r)+448 m (r-2 m)^2}{4 r (r-2 m)^2} 
	- \frac{r_0^2 (104 m-11 r_0)+448 m (r_0-2 m)^2}{4 r_0 (r_0-2 m)^2}  \notag \\
	&\quad + 8\Bigg(\frac{(17m^2 - 24mr + 8r^2)\sqrt{1-\frac{2m}{r}}}{m(r-2m)^2} -\frac{(17m^2 - 24mr_0 + 8r_0^2)\sqrt{1-\frac{2m}{r_0}}}{m(r_0-2m)^2} \Bigg)
	- \frac{127}{2}\ln\left(\frac{r_0(r-2m)}{r(r_0-2m)}\right) \bigg\}. \label{eq:4.84'}
\end{align}
The commutative limits of the affine parameter and coordinate time are recovered by setting $\Theta = 0$ \cite{Chandr1}.

\begin{figure}[h]
	\begin{center}
		\includegraphics[width=0.5\textwidth]{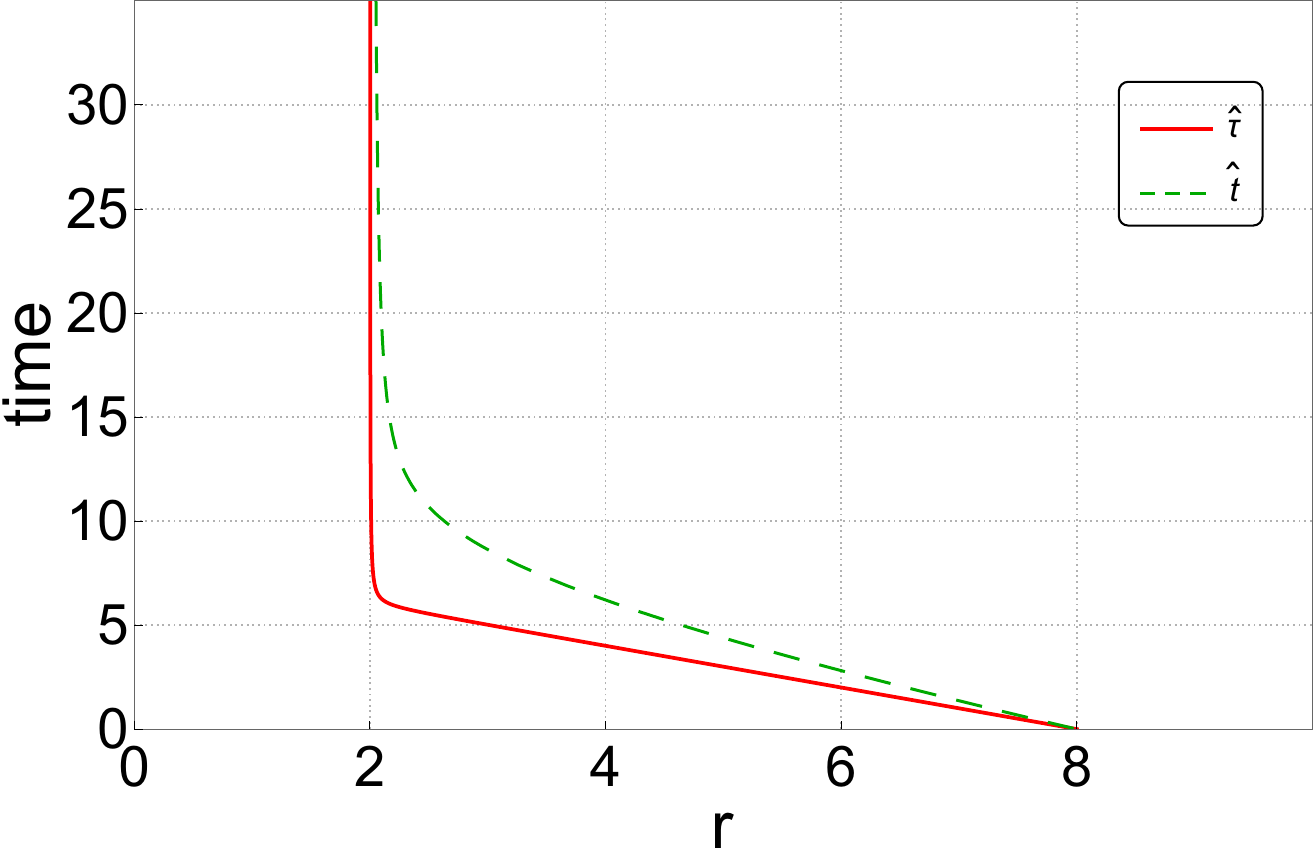}
	\end{center}
	\caption{Variation of the affine parameter $\hat{\tau}$ and coordinate time $\hat{t}$ along a radial null geodesic of a photon in NC Schwarzschild spacetime as a function of $r$.}
	\label{figd}
\end{figure}

Fig. \ref{figd} illustrates the behavior of $\hat{\tau}$ and $\hat{t}$ for a photon initially at $r = r_0$ falling toward the NC Schwarzschild BH. Notably, in the affine parameter framework of NC spacetime, photons require an infinite affine parameter to reach the NC singularity, in contrast to the commutative case \cite{Chandr1}. The same observation applies to coordinate time, where the photon cannot reach the singularity within a finite time. This behavior aligns with the motion of massive particles and highlights the role of non-commutativity in preventing singularity encounters within a finite temporal or affine framework. Additionally, for massless particles, our analysis predicts a novel free-fall behavior in NC spacetime using the NC gauge theory of gravity. This differs from the NC matter distribution approach in Refs. \cite{rahaman,bhar}, which modifies the commutative behavior and alters the underlying physics of the spacetime geometry. In contrast, non-commutativity should act as a quantum correction to the metric, reducing to the commutative case when this correction vanishes. While an NC matter distribution can recover the commutative free-fall motion, as demonstrated in Ref. \cite{kuniyal}, it does not introduce new predictions for free-fall dynamics. However, our approach, based on the NC gauge theory of gravity, preserves the commutative solution while predicting novel macroscopic effects of non-commutativity.


\subsection{Circular motion of massless particles}

This section explores the null geodesic motion of photons in the NC Schwarzschild black hole (BH) spacetime. The effective potential for a photon is derived from \eqref{eq:4.7} by setting $h = 0$, $c = 1$, and $l \neq 0$.
\begin{figure}[h]
	\begin{center}
		
		\includegraphics[width=0.5\textwidth]{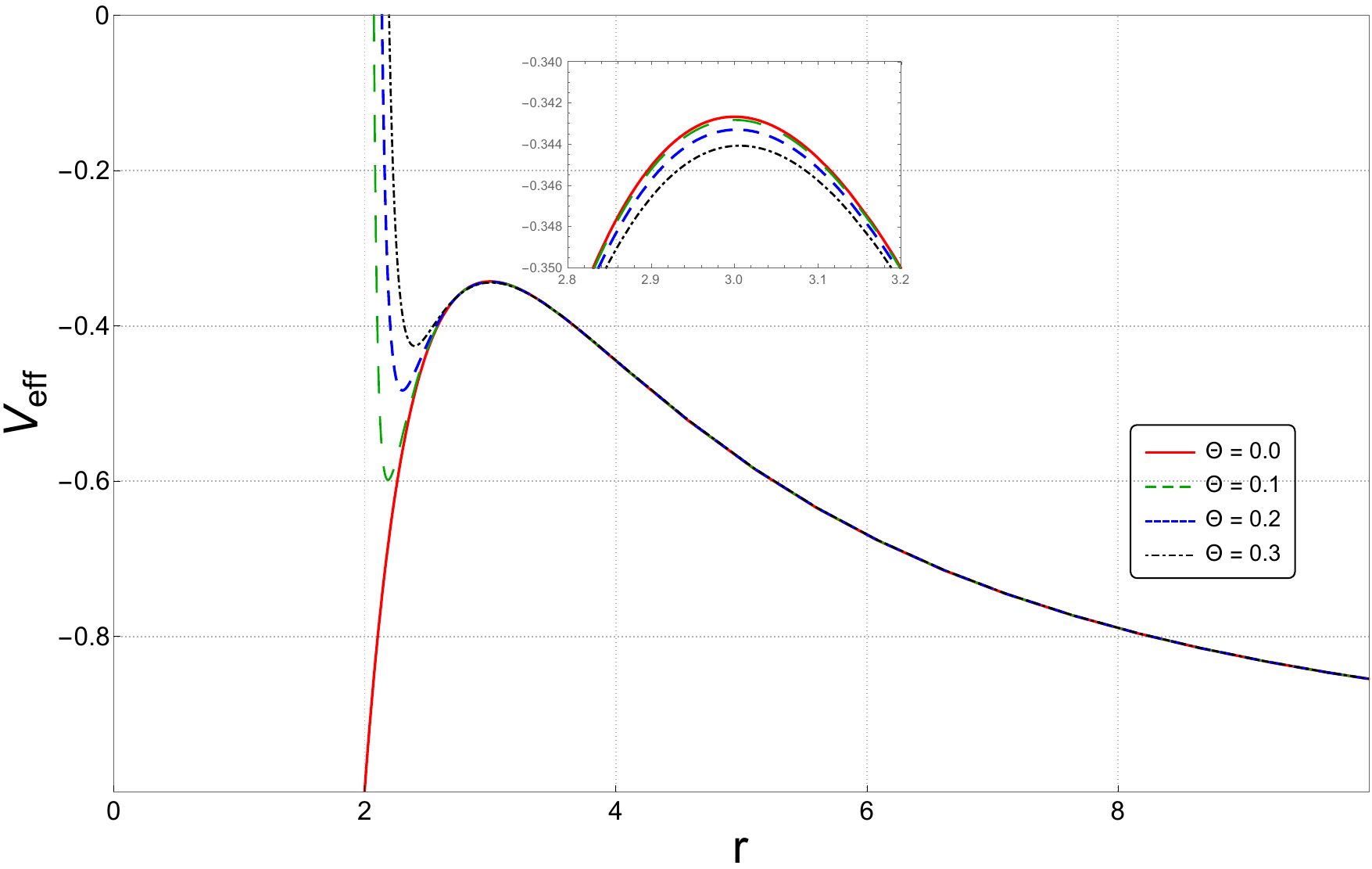}\hfill
		\includegraphics[width=0.5\textwidth]{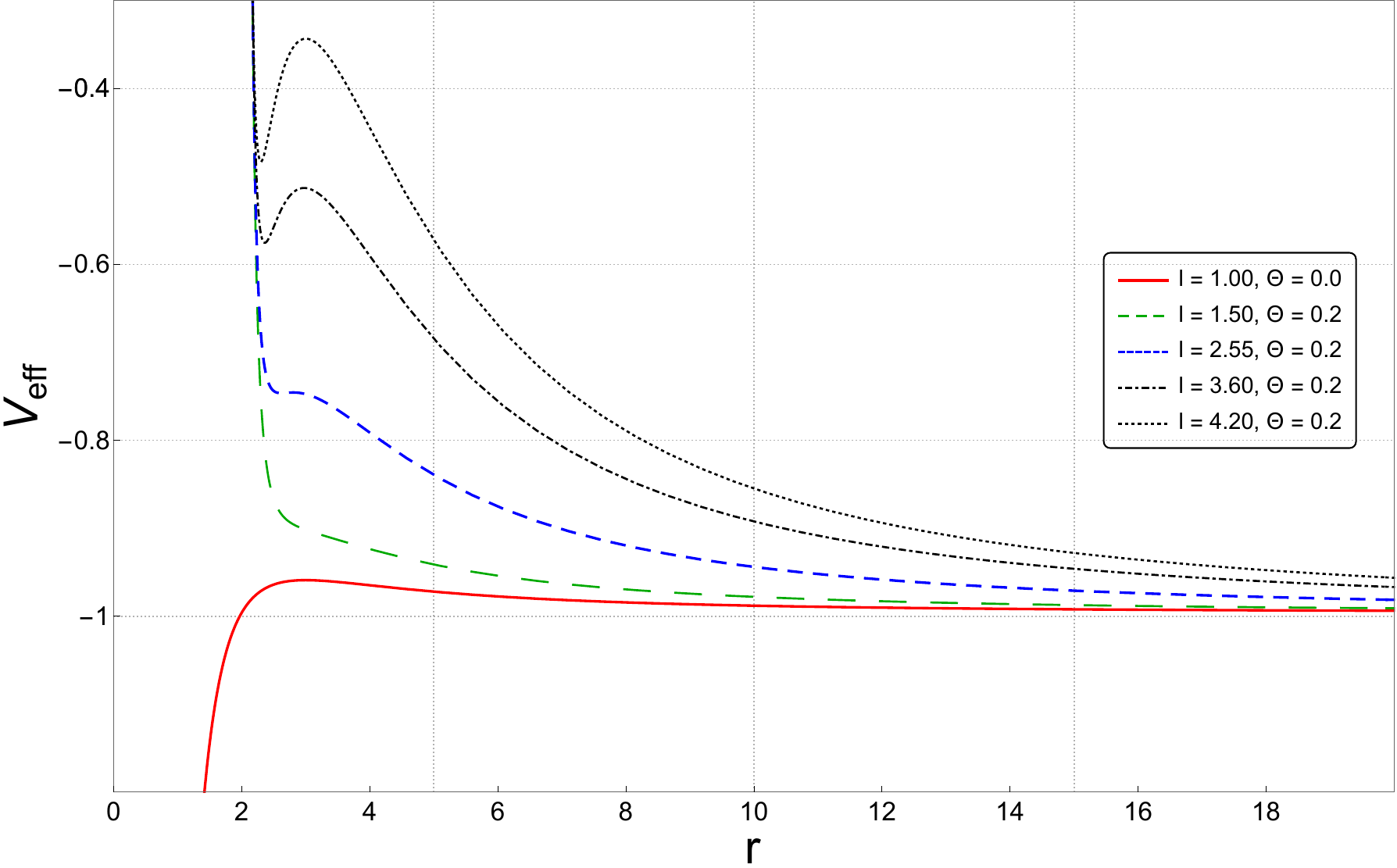}
	\end{center}
	\caption{Behavior of the effective potential for a massless particle with varying non-commutative parameter $\Theta$ and fixed values of $E=0.998$, $m=1$, $l=4.2$ (left panel), and with varying angular momentum $l$ and fixed $E=0.998$, $m=1$, $\Theta=0.2$ (right panel).}
	\label{fig2.1}
\end{figure}

Fig. \ref{fig2.1} illustrates the influence of the NC parameter $\Theta$ (left panel) and the angular momentum $l$ (right panel) on the effective potential of a massless test particle (photon). As shown in the left panel, introducing non-commutativity leads to the emergence of a new minimum in the effective potential near the event horizon. This phenomenon is analogous to that observed for massive test particles (see SubSec.~\ref{subsec:NCCM}). However, for massless particles, only two extrema exist outside the event horizon, allowing the interpretation of the new minimum as a stable circular orbit. Consequently, this results in the formation of a stable photon sphere in NC spacetime. Additionally, the position of this new minimum shifts farther from the event horizon as $\Theta$ increases, while NC effects become negligible at large distances ($r \gg r_h^{NC}$). The right panel of Fig.~\ref{fig2.1} reveals that, in NC spacetime, the effective potential exhibits extremal points only when the angular momentum satisfies a specific condition. This contrasts with the commutative case, where the effective potential always possesses a maximum, regardless of the angular momentum. For $\Theta = 0.2$, the effective potential exhibits two extrema when the angular momentum exceeds a critical value, $l^{\text{crit}} \geq 2.55$. These critical values depend on the NC parameter $\Theta$. In conclusion, NC spacetime consistently supports the existence of two photon spheres: an inner stable photon sphere and an outer unstable photon sphere. This behavior significantly differs from the commutative case and highlights the unique role of NC geometry in modifying the geodesic structure of spacetime. Furthermore, these results are novel and distinct from previous works using different NC approaches such as NC matter distributions \cite{kuniyal,bhar,rahaman,NCLD2}, NC gauge theory of gravity via deformed geometry \cite{NCGTG2,NCGTG3}, or deformed mass in gauge theory \cite{NCGTGmassdeformed1,NCGTGmassdeformed2,NCGTGmassdeformed3}, which did not predict the existence of a stable photon sphere near the event horizon.

\subsubsection{Null circular orbit and photon sphere}

The conditions for circular photon orbits in NC Schwarzschild spacetime are analogous to those for massive particles, as given in Eq.~\eqref{eq:27}. By utilizing the effective potential \eqref{eq:4.7} with $h = 0$, the relationship between the photon’s energy $E_c$ and angular momentum $L_c$ can be expressed as:

\begin{equation}\label{eq:impactparameter}
	\frac{1}{D_c}=\frac{E_c}{L_c}=\sqrt{\frac{r_c-2m}{r_c^3}+\left(\frac{J(r_c)+G(r_c)\sqrt{1-\frac{2m}{r_c}}}{8 r_c^6 (r_c -2m)}\right)\Theta^2},
\end{equation}
where $D_c$ is the impact parameter, and the functions $J(r_c)$ and $G(r_c)$ are defined as:
\begin{equation}
	J(r_c)=152 m^3-161 m^2r_c+52 mr_c^2-5 r_c^3, \quad G(r_c)=29 m^2r_c-27 mr_c^2+3 r_c^3.
\end{equation}

\begin{table}[h]
	\begin{center}
		\caption{Numerical values for the unstable $r_{\text{uns}}$ and the new stable circular orbit $r_{\text{sta}}$ of the photon sphere in NC spacetime for different $\Theta$ values, with $E=0.998$, $l=4$, and $m=1$.}
		\label{tab1}
		\begin{tabular}{ c c c c c c c }
			\hline
			$\Theta$	& 0 & 0.10 & 0.15 & 0.20 & 0.25 & 0.30 \\
			\hline
			$r_{\text{sta}}$ (new)	&  & 2.19378 & 2.25867 & 2.31807 & 2.37379 & 2.42689 \\
			$r_{\text{uns}}$	& 3.00000 & 2.99886 & 2.99739 & 2.99528 & 2.99246 & 2.98881 \\
			\hline
		\end{tabular}
	\end{center}
\end{table}

Table \ref{tab1} presents numerical solutions of Eq.~\eqref{eq:27}, illustrating how the stable and unstable photon sphere radii depend on $\Theta$. For $3.17 \leq l \leq 4.12$ (where this range corresponds to $0 \leq \Theta \leq 0.3$ and energy $E=1$), the stable circular orbit radius increases with $\Theta$, while the unstable orbit radius decreases. However, for $l > 4.12$, both orbit radii increase as $\Theta$ grows.
\begin{figure}[h]
	\begin{center}
		\includegraphics[width=0.25\textwidth]{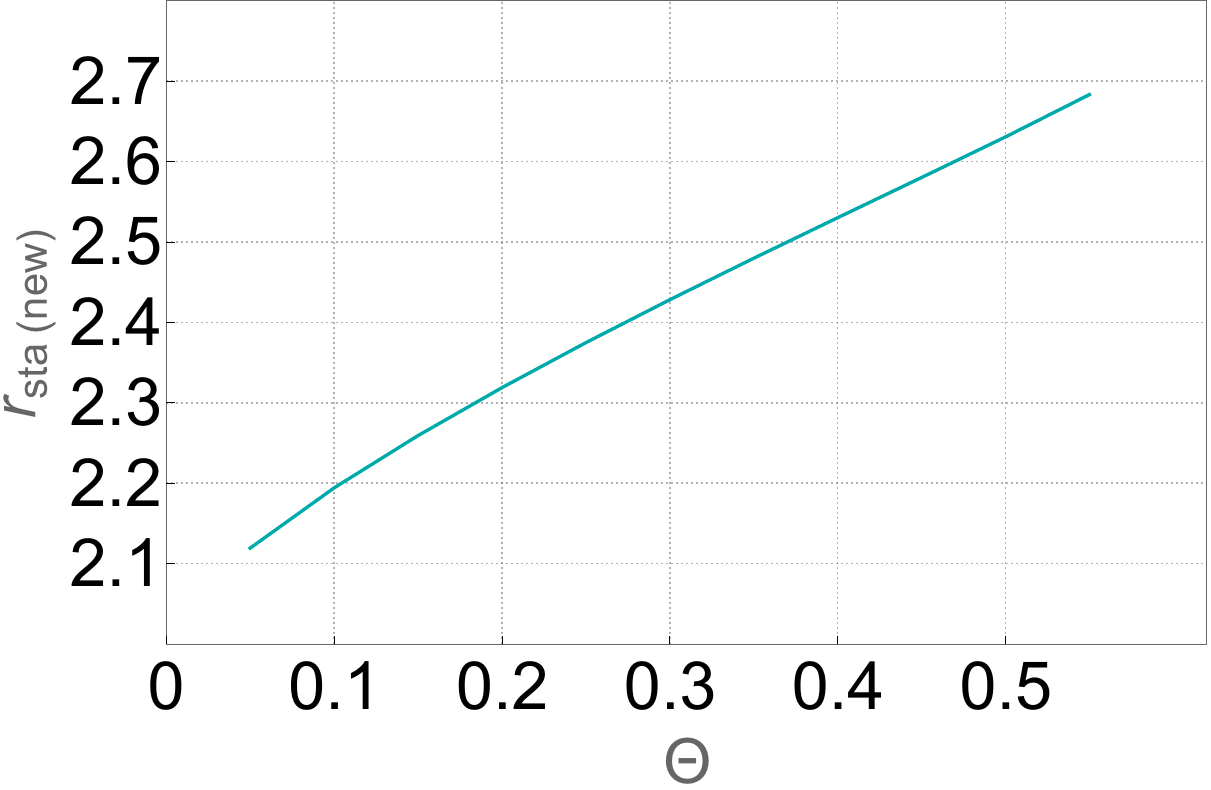}\hfill
		\includegraphics[width=0.25\textwidth]{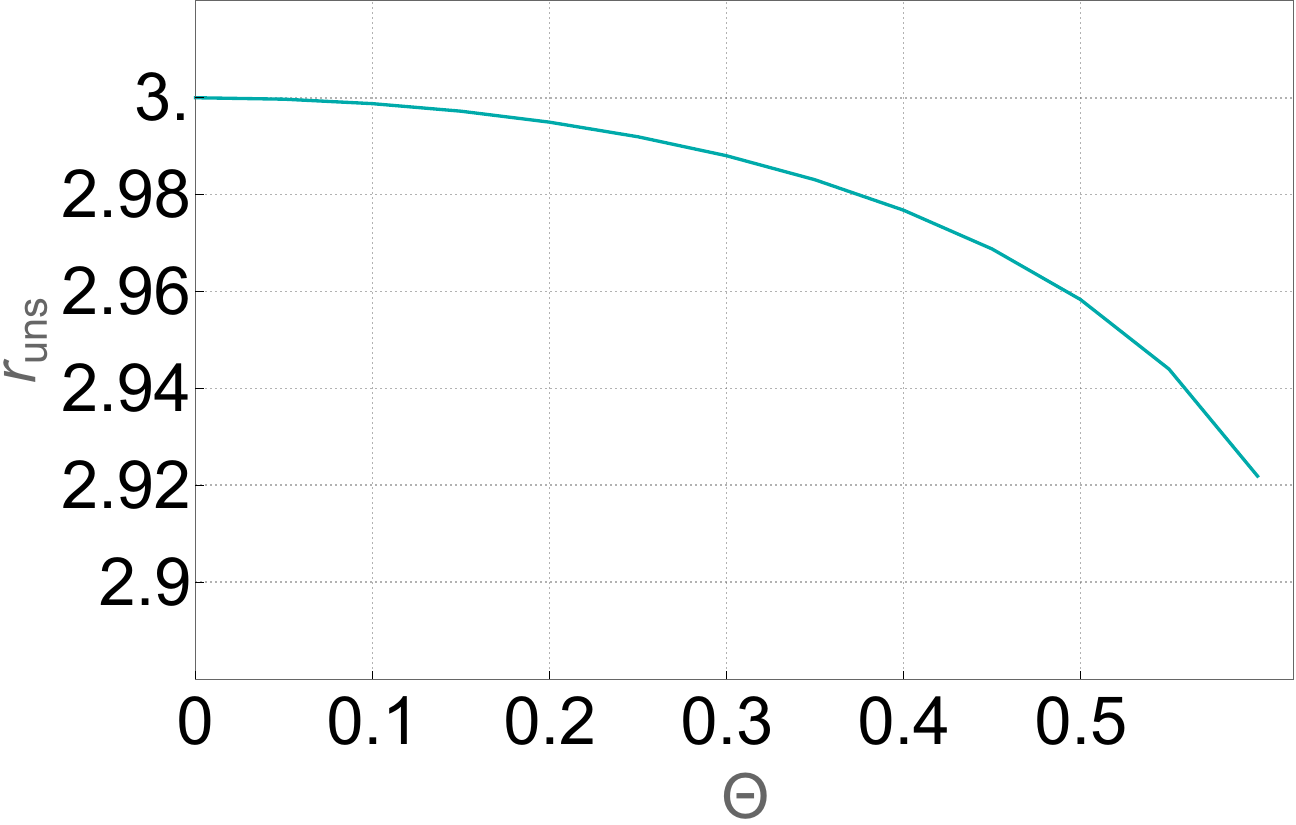}\hfill
		\includegraphics[width=0.25\textwidth]{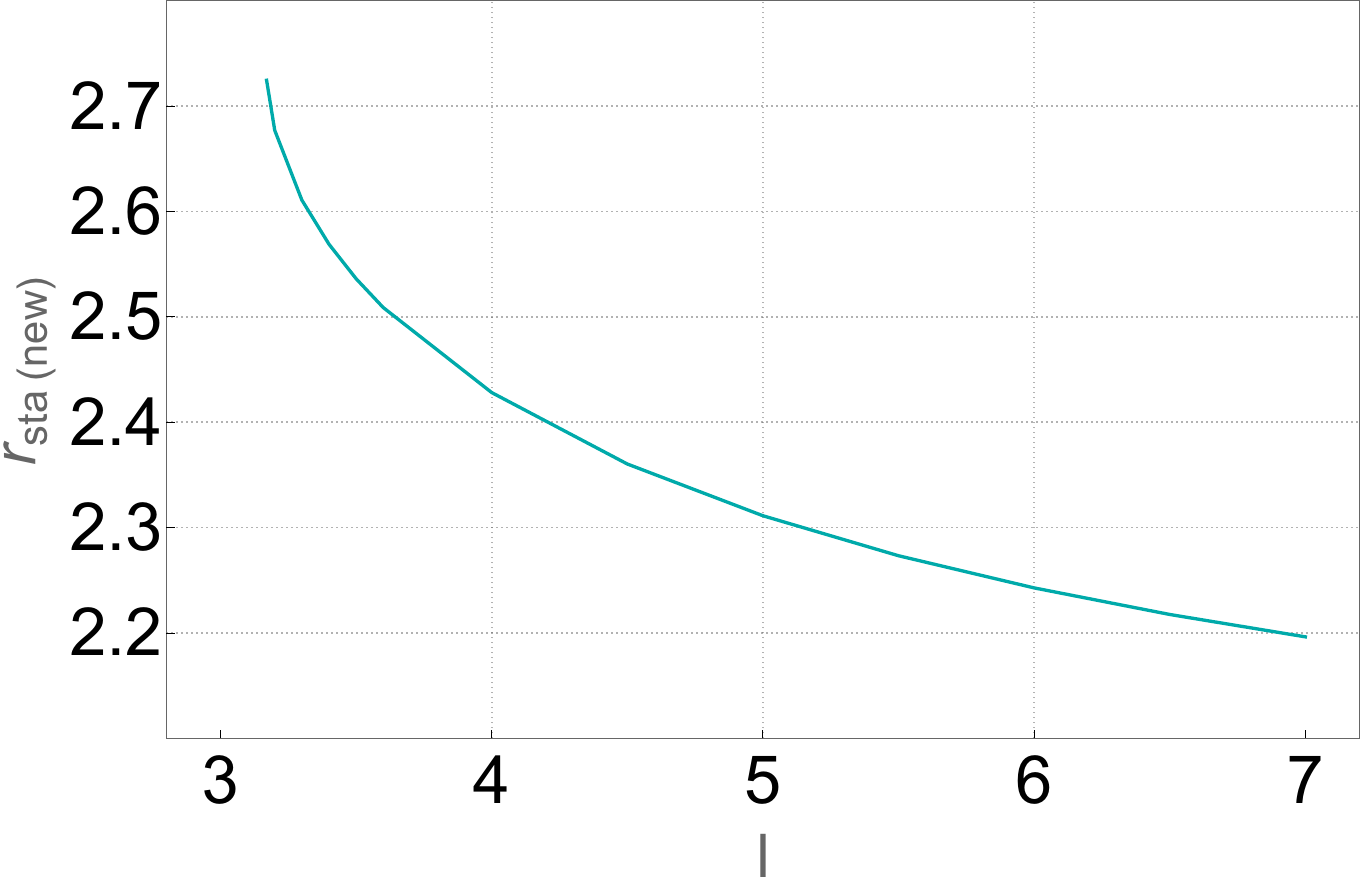}\hfill
		\includegraphics[width=0.25\textwidth]{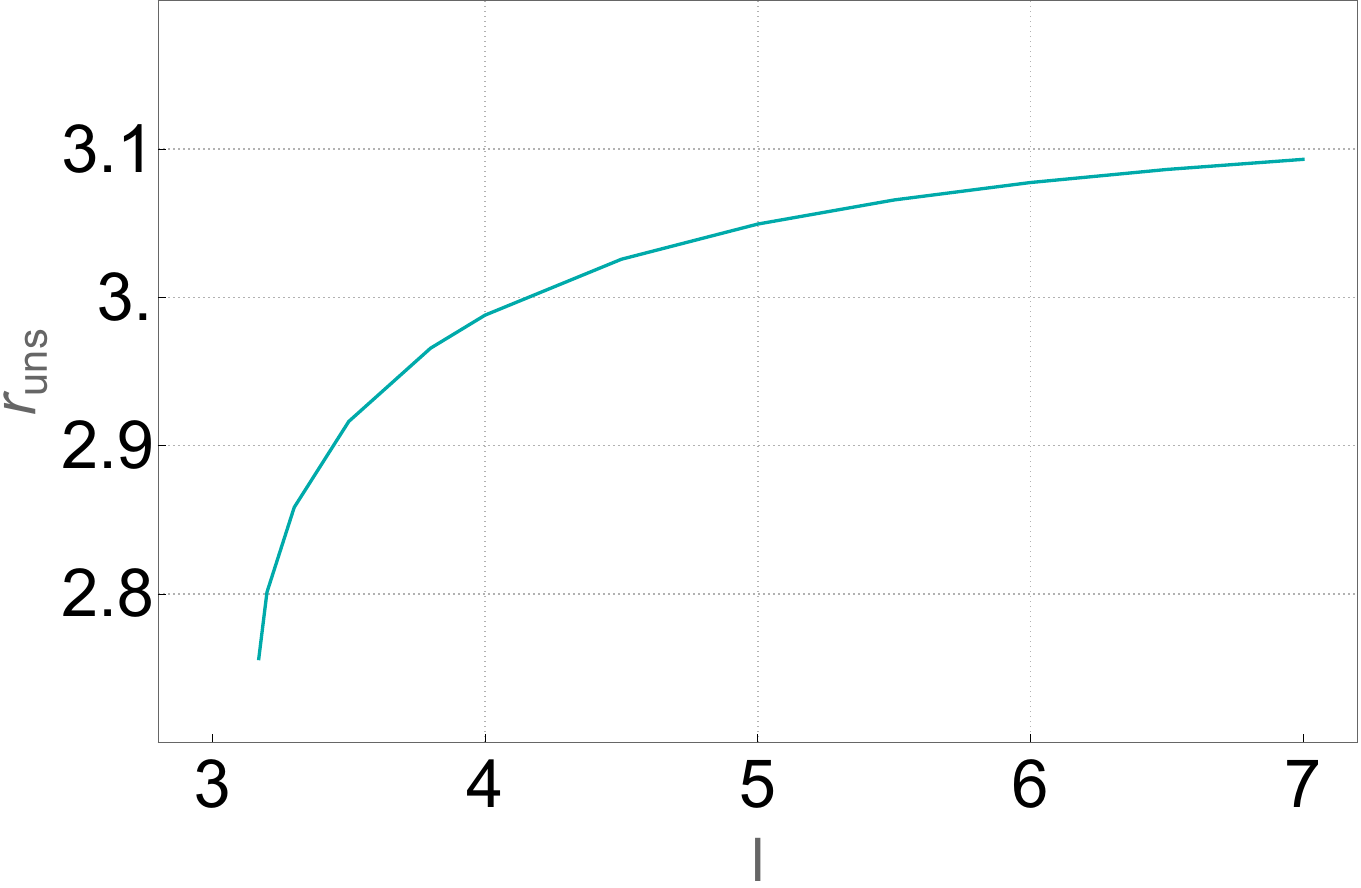}
	\end{center}
	\caption{Behavior of the photon sphere radius in NC spacetime. Stable and unstable circular orbits are shown as functions of $\Theta$ for fixed $l=4$, $E=1$, $m=1$ (two left panels), and as functions of $l$ for fixed $\Theta=0.3$, $E=1$, $m=1$ (two right panels).}
	\label{fig2.2}
\end{figure}

The left panels of Fig.~\ref{fig2.2} illustrate that as $\Theta$ increases, the radius of the stable circular orbit grows, while the radius of the unstable orbit decreases. In the NC framework, the unstable photon orbit has a smaller radius than in the commutative case for $l < 4.12$, whereas for $l \geq 4.12$, its radius exceeds the commutative value. This divergence arises from modifications to the gravitational field induced by non-commutativity. The right panels of Fig.~\ref{fig2.2} show the dependence of circular orbit radii on angular momentum $l$. For a fixed $\Theta = 0.3$, the stable circular orbit radius decreases as $l$ increases, while the unstable orbit radius grows. This behavior highlights the interplay between angular momentum and the NC parameter in shaping the geodesic structure of spacetime.

	\begin{figure}[h]
		\centering
		\begin{tikzpicture}[scale=0.8]
			\draw (0,0) circle (2.0675cm);
			\filldraw[color=black!100, fill=gray!80, very thick] (0,0) circle (2cm); 
			\draw [ blue!100, very thick] (0,0) circle (2.35657cm); 
			\draw [ orange!100, very thick] (0,0) circle (3.06212cm); 	
			\node[above] at (0,0) {\text{\small NC SBH }} ;singularity
			\node[below] at (0,0) {\text{\small Singularity }} ;
			\draw[dashed,black, very thick] (0,-2.0675) -- (6,-2.0675)
			node[above left]{NC event horizon} ;
			\draw[dashed,blue, very thick] (0,2.35657) -- (-6,2.35657)
			node[above right]{Stable photon sphere};
			\draw[dashed,orange!100, very thick] (0,-3.06212) -- (-6,-3.06212)
			node[above right]{Unstable photon sphere};
		\end{tikzpicture}	
		
		\caption{The schematic picture of the stable/unstable photon sphere around the NC SBH.}
		\label{fig2.4}
	\end{figure}
	\begin{tikzpicture}
	\end{tikzpicture}

\subsubsection{Lyapunov exponents}

In the following, we analyze the stability and instability of the two photon spheres in NC spacetime using the Lyapunov exponent. Since proper time does not exist for null circular orbits, only the coordinate time Lyapunov exponent is considered.

We begin by computing the angular frequency and the coordinate time period, given by Eqs.~\eqref{eq:omega} and \eqref{eq:period}, respectively:
\begin{equation}
	\hat{\Omega}_c^{\text{Null}} = \frac{1}{D_c}, \quad \hat{T}_{\hat{\Omega}}^{\text{Null}} = 2\pi D_c.
\end{equation}

Next, we derive the coordinate time Lyapunov exponent for photon orbits around the NC Schwarzschild BH as:

\begin{equation}
	\hat{\lambda}_c^{\text{Null}} = \sqrt{-\frac{3((r_c-4m )(r_c-2m ))}{r_c^4} 
		+ \left(\frac{O(r_c) + I(r_c)\sqrt{1 - \frac{2m}{r_c}}}{16(r_c - 2m)^2 r_c^7}\right)\Theta^2},
\end{equation}
where
\begin{subequations}
	\begin{align}
		O(r_c) &= -19008 m^5 + 36600 m^4r_c - 26234 m^3r_c^2 + 8463 m^2r_c^3 - 1236 mr_c^4 + 70 r_c^5, \\
		I(r_c) &= -2791 m^4r_c + 4791 m^3r_c^2 - 2999 m^2r_c^3 + 738 mr_c^4 - 42 r_c^5.
	\end{align}
\end{subequations}

\begin{figure}[h]
	\begin{center}
		\includegraphics[width=0.5\textwidth]{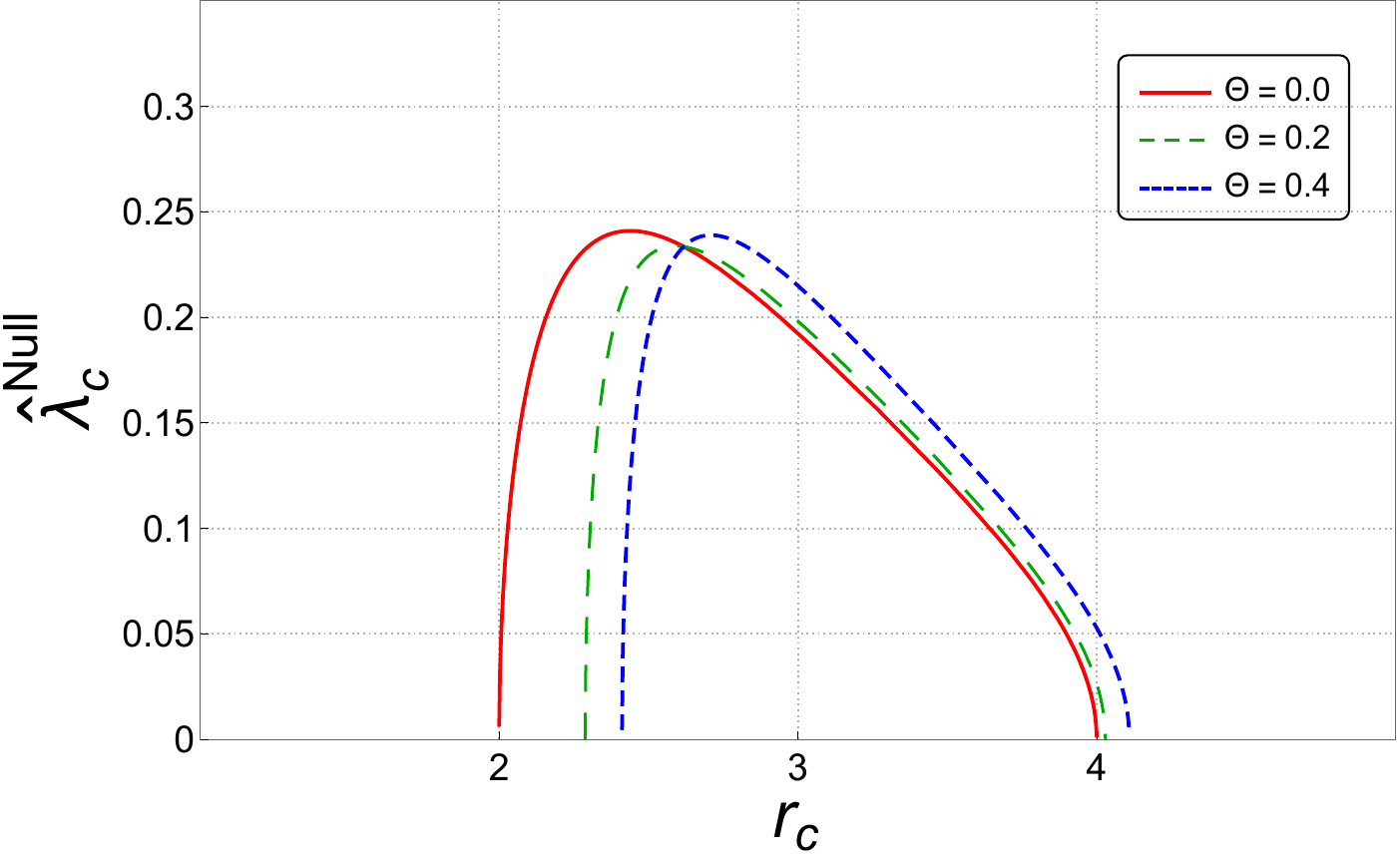}\hfill
		\includegraphics[width=0.49\textwidth]{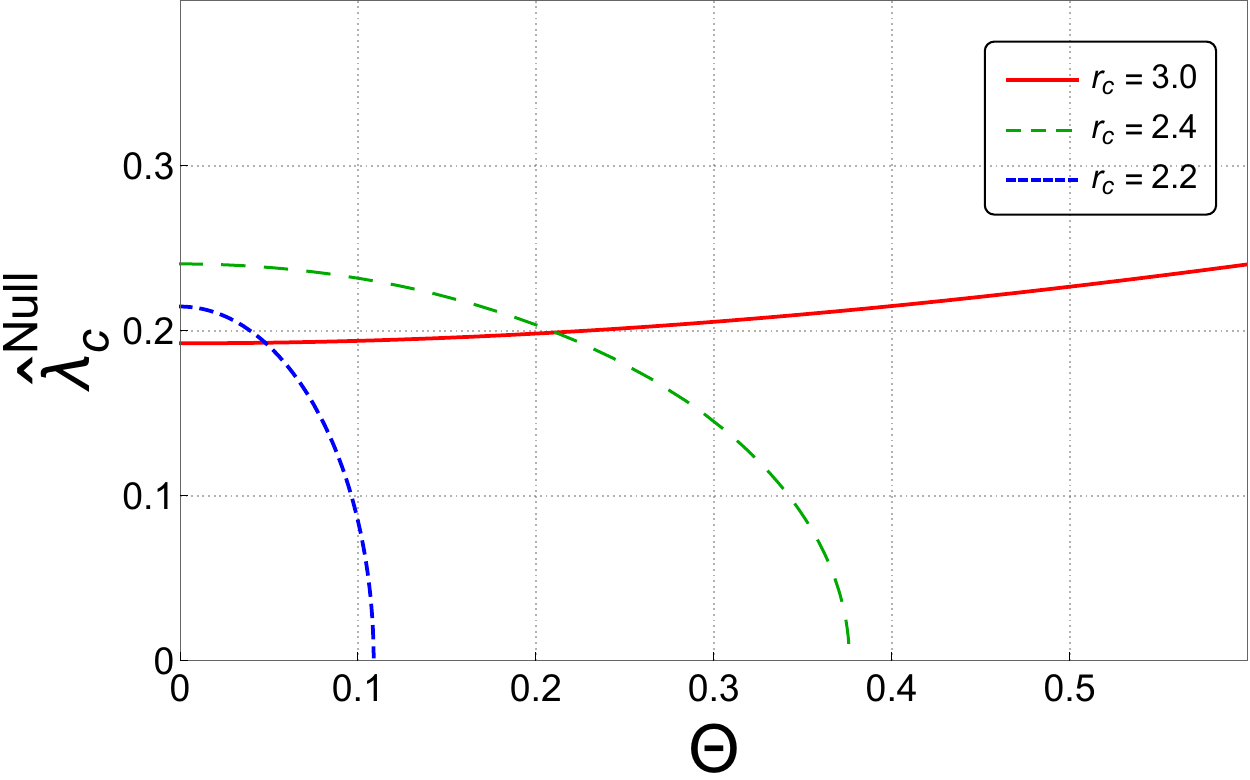}
	\end{center}
	\caption{Coordinate time Lyapunov exponent for massless particles as a function of $r_c$ (left panel) and the NC parameter $\Theta$ (right panel).}
	\label{fig2.5}
\end{figure}

Fig. \ref{fig2.5} illustrates the behavior of the coordinate time Lyapunov exponent for photons as a function of $r_c$ (left panel) and $\Theta$ (right panel). The left panel shows that instability initially increases to a peak whose magnitude increases with higher $\Theta$, before decreasing as $r_c$ increases, eventually vanishing for $r_c \geq 4$. In NC spacetime, orbits near the event horizon exhibit reduced instability compared to the commutative case, whereas at larger distances, this trend reverses. Moreover, while instability in commutative spacetime begins at the event horizon, NC geometry shifts it outward, allowing for stable orbits to emerge near the event horizon. This observation aligns with the effective potential analysis for massless test particles, confirming the existence of a stable photon sphere around the NC Schwarzschild BH. Similar behavior has been reported in other NC geometries, as discussed in Ref.~\cite{geostab5}, where stable photon spheres cannot form within the event horizon. The right panel examines the dependence of circular orbit radii $r_c$ on $\Theta$. It reveals that the orbit at $r_c = 3$ remains unstable in NC spacetime, with instability increasing as $\Theta$ grows. In contrast, the newly emerging circular orbits near the event horizon exhibit reduced instability, which further decreases with increasing $\Theta$, eventually leading to stabilization. This suggests the existence of stable circular orbits near the event horizon in NC spacetime.

The ratio of the Lyapunov exponent to the angular frequency in coordinate time, $\frac{\hat{\lambda}_c^{\text{Null}}}{\hat{\Omega}_c^{\text{Null}}}$, defines the instability exponent of null circular orbits in NC spacetime:

\begin{equation}
	\frac{\hat{\lambda}_c^{\text{Null}}}{\hat{\Omega}_c^{\text{Null}}} = \sqrt{\frac{3(4m - r_c)}{r_c} + \left(\frac{O(r_c) + 6J(r_c)(8m^2 - 6mr_c + r_c^2) 
			+ (I(r_c) + 6G(r_c)(8m^2 - 6mr_c + r_c^2))\sqrt{1 - \frac{2m}{r_c}}}{16(r_c - 2m)^3 r_c^4}\right)\Theta^2}.
\end{equation}

\begin{figure}[h]
	\begin{center}
		\includegraphics[width=0.6\textwidth]{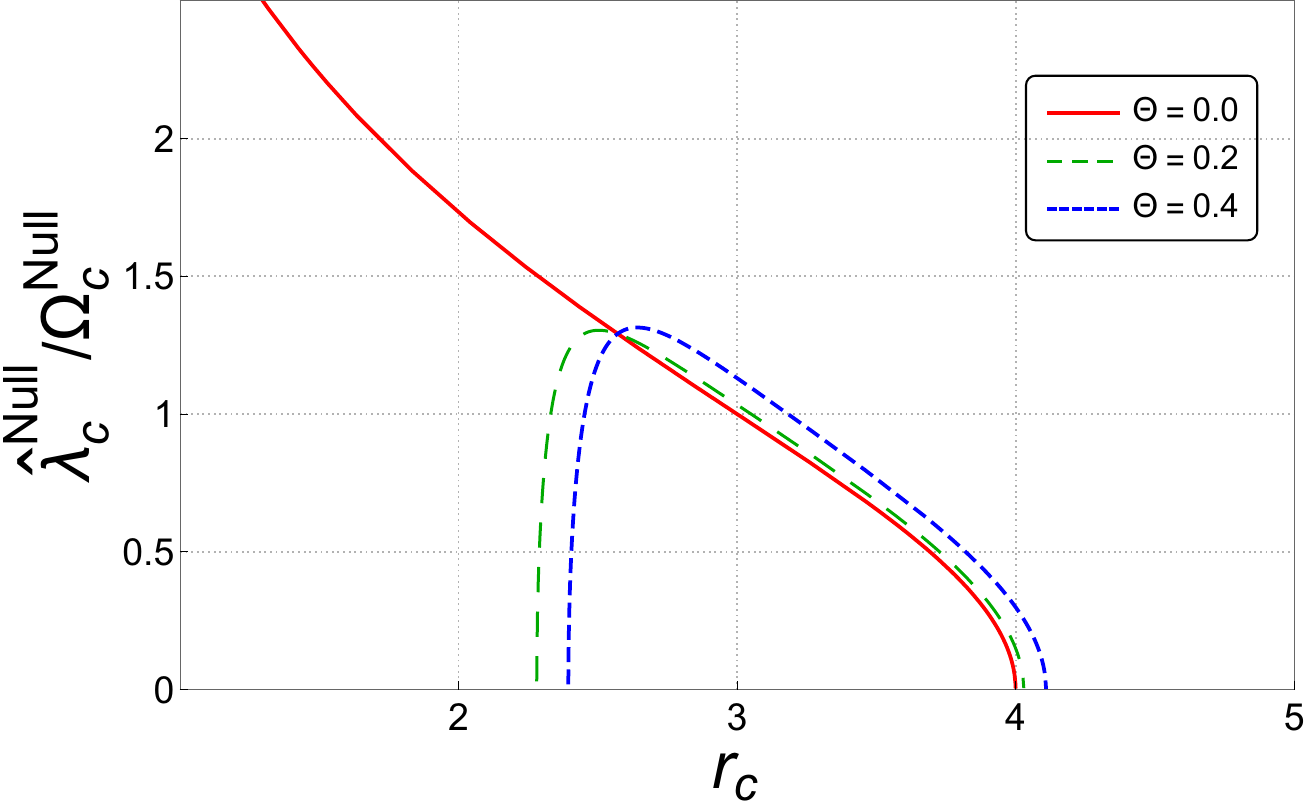}
	\end{center}
	\caption{Ratio of the Lyapunov exponent to the angular frequency in coordinate time as a function of $r_c$.}
	\label{fig2.6}
\end{figure}

Fig. \ref{fig2.6} illustrates the variation of $\hat{\lambda}_c^{\text{Null}}/\hat{\Omega}_c^{\text{Null}}$ with $r_c$, exhibiting behavior similar to that in Fig. \ref{fig2.5}. Unlike in the commutative case, NC spacetime permits the existence of stable orbits near the event horizon. This contrasts with the findings of Ref.~\cite{geostab5}, which confirm the stability of the newly emerging photon sphere.

Finally, the critical exponent $\hat{\gamma}$ for null circular orbits is defined as the ratio of the Lyapunov timescale, $\hat{T}_{\hat{\lambda}} = \frac{1}{\hat{\lambda}}$, to the orbital timescale, $\hat{T}_{\hat{\Omega}} = \frac{2\pi}{\hat{\Omega}}$ \cite{geostab11,geostab3,geostab12}:
\begin{equation}\label{eq:4.55}
	\hat{\gamma}^{\text{Null}} = \frac{\hat{T}_{\hat{\lambda}}^{\text{Null}}}{\hat{T}_{\hat{\Omega}}^{\text{Null}}} = 
	\frac{1}{2\pi \sqrt{\frac{3(4m - r_c)}{r_c} + \left(\frac{O(r_c) + 6J(r_c)(8m^2 - 6mr_c + r_c^2) + (I(r_c) + 6G(r_c)(8m^2 - 6mr_c + r_c^2))\sqrt{1 - \frac{2m}{r_c}}}{16(r_c - 2m)^3 r_c^4}\right)\Theta^2}}.
\end{equation}
In the limit $\Theta = 0$, we recover the commutative case.
\begin{table}[h]
	\begin{center}
		\caption{Critical exponent $\hat{\gamma}^{\text{Null}}$ for unstable orbits $r_{\text{uns}}$ and stable orbits $r_{\text{sta}}$ of the photon sphere in NC spacetime for different values of $\Theta$, with $m=1$.}
		\label{tab2}
		\begin{tabular}{c c | c c c c }
			\hline
			\hline
			& $\Theta$  & $r_c=3$ & $r_c=2.4$ & $r_c=2.3$ & $r_c=2.2$ \\
			\hline$\hat{\gamma}^{\text{Null}}$ 
			& 0.0  & 0.159155 & -- & -- & -- \\
			& 0.1  & 0.157785 & 0.115961 & 0.119163 & 0.228298 \\
			& 0.2  & 0.153879 & 0.128464 & 0.228944		    & $0 - 0.0683652i$ \\
			& 0.3  & 0.147966 & 0.162989 & $0 - 0.1226300i$ & $0 - 0.0407389i$ \\
			& 0.4  & 0.140728 & 0.426016 & $0 - 0.0732637i$ & $0 - 0.0295329i$ \\
			\hline
			\hline
		\end{tabular}
	\end{center}
\end{table}

The numerical values of the critical exponent $\hat{\gamma}^{\text{Null}}$ for photons in various circular orbits $r_c$ and different values of the NC parameter $\Theta$ are summarized in Table~\ref{tab2}. This table highlights the presence of both real and imaginary values of $\hat{\gamma}^{\text{Null}}$. Real positive values indicate unstable circular orbits \cite{geostab11}, whereas imaginary values correspond to stable circular orbits. For the photon sphere orbit in the Schwarzschild BH spacetime at $r_c = 3$, instability persists even in the NC framework. According to the definition in Eq.~\eqref{eq:4.55}, this implies that the Lyapunov timescale $\hat{T}_{\hat{\lambda}}$ is shorter than the orbital timescale $\hat{T}_{\hat{\Omega}}$, i.e., $\hat{T}_{\hat{\lambda}} < \hat{T}_{\hat{\Omega}}$ \cite{geostab12}. A similar behavior is observed for other values of $r_c$ in Table~\ref{tab2}, where real values of $\hat{\gamma}^{\text{Null}}$ consistently signify instability. Furthermore, the impact of the NC parameter $\Theta$ on $\hat{\gamma}^{\text{Null}}$ is evident. As $\Theta$ increases, the values of $\hat{\gamma}^{\text{Null}}$ decrease for $r_c = 3$, indicating increasing instability. The smallest values of $\hat{\gamma}^{\text{Null}}$ correspond to stronger Lyapunov instability \cite{geostab3} (see the right panel of Fig.~\ref{fig2.5}). On the other hand, the imaginary values of $\hat{\gamma}^{\text{Null}}$, which denote stable circular orbits, align with the new minima in the effective potential observed for $l = 4.2$ (values comparable to those in Table~\ref{tab1}). This demonstrates the emergence of a new stable photon sphere in NC spacetime, contrasting with the instability of the external photon sphere in both commutative and NC scenarios.

\subsubsection{Shadow radius and limit of the non-commutativity parameter}

In the equatorial plane ($\theta = \pi/2$), the radius of the BH shadow, $\hat{R}_{\text{shadow}}$, is directly related to the impact parameter 
$D_c = \bigg|\frac{L_c}{E_c}\bigg|$ by the relation $\hat{R}_{\text{shadow}} = D_c\big|_{r = r^{\text{uns}}_{\text{ps}}}$. This is significant because a light beam emitted from infinity with the critical impact parameter $D_c = D_c\big|_{r = r^{\text{uns}}_{\text{ps}}}$ will precisely reach the unstable circular orbit of the photon, while a light beam emitted with an impact parameter $D_c < D_c\big|_{r = r^{\text{uns}}_{\text{ps}}}$ will fall into the BH. The radius $r = r^{\text{uns}}_{\text{ps}}$ represents the location of the unstable photon sphere \cite{BHShadow1,Hassanabadi2}. This radius can be determined by solving the equation $\frac{dV_{\text{eff}}}{dr} = 0$. To leading order in the NC parameter $\Theta$ and the BH mass $m$, the solution is given by:

\begin{equation}\label{eq:4.56}
	\hat{r}^{\text{uns}}_{\text{ps}} = 3m + \left(\frac{15 + 74 \sqrt{3}}{216m}\right)\Theta^2\,.
\end{equation}
It is important to note that the radius of the stable photon sphere in NC spacetime can only be obtained numerically (see Table~\ref{tab1}). Using the expression \eqref{eq:4.56} together with \eqref{eq:impactparameter}, the shadow radius of the NC Schwarzschild BH can be expressed using $\hat{R}_{\text{shadow}} = D_c\big|_{r = r^{\text{uns}}_{\text{ps}}}$:
\begin{equation}\label{eq:4.57}
	\hat{R}_{\text{shadow}}=\Bigg|\frac{L_c}{E_c}\Bigg|_{r=r^{\text{uns}}_{\text{ps}}} = 3\sqrt{3}m\left(1 + \frac{(25\sqrt{3}-2)}{432m^2}\Theta^2\right)\,,
\end{equation}
where the corresponding commutative case is recovered by setting $\Theta = 0$.

\begin{figure}[h]
	\begin{center}
		\includegraphics[width=0.495\textwidth]{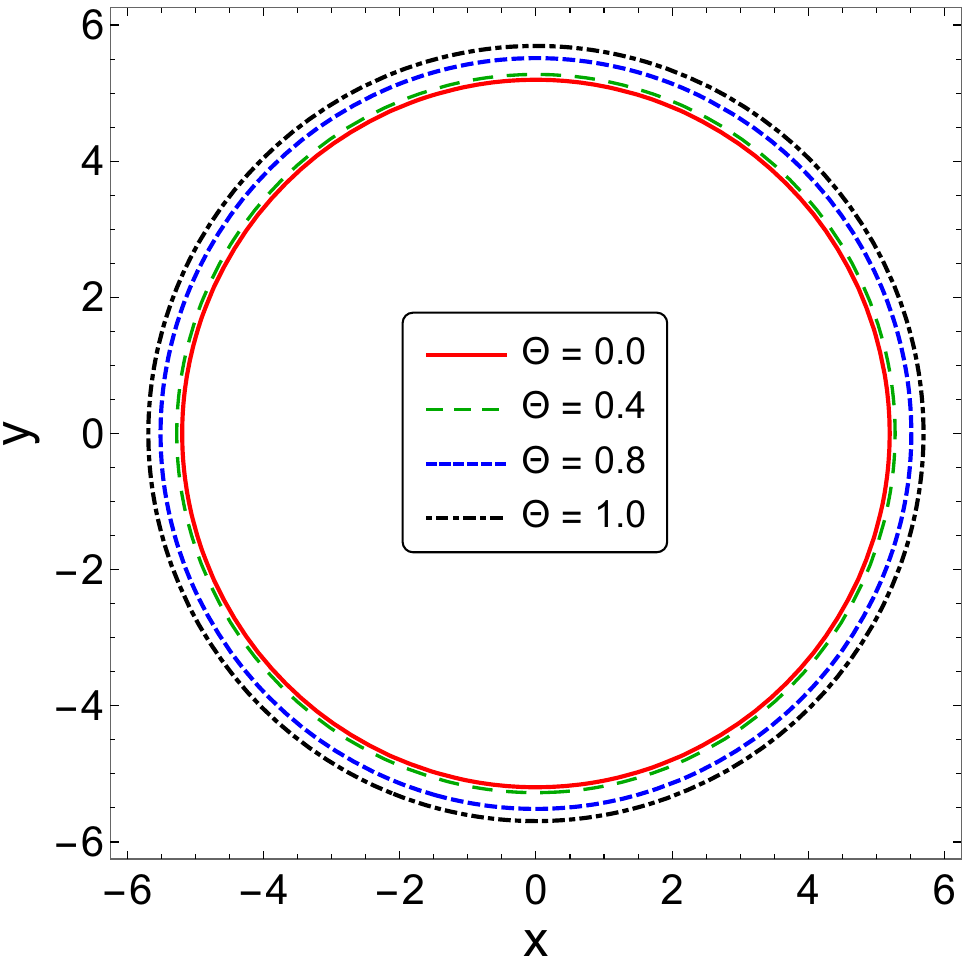}\hfill
		\includegraphics[width=0.495\textwidth]{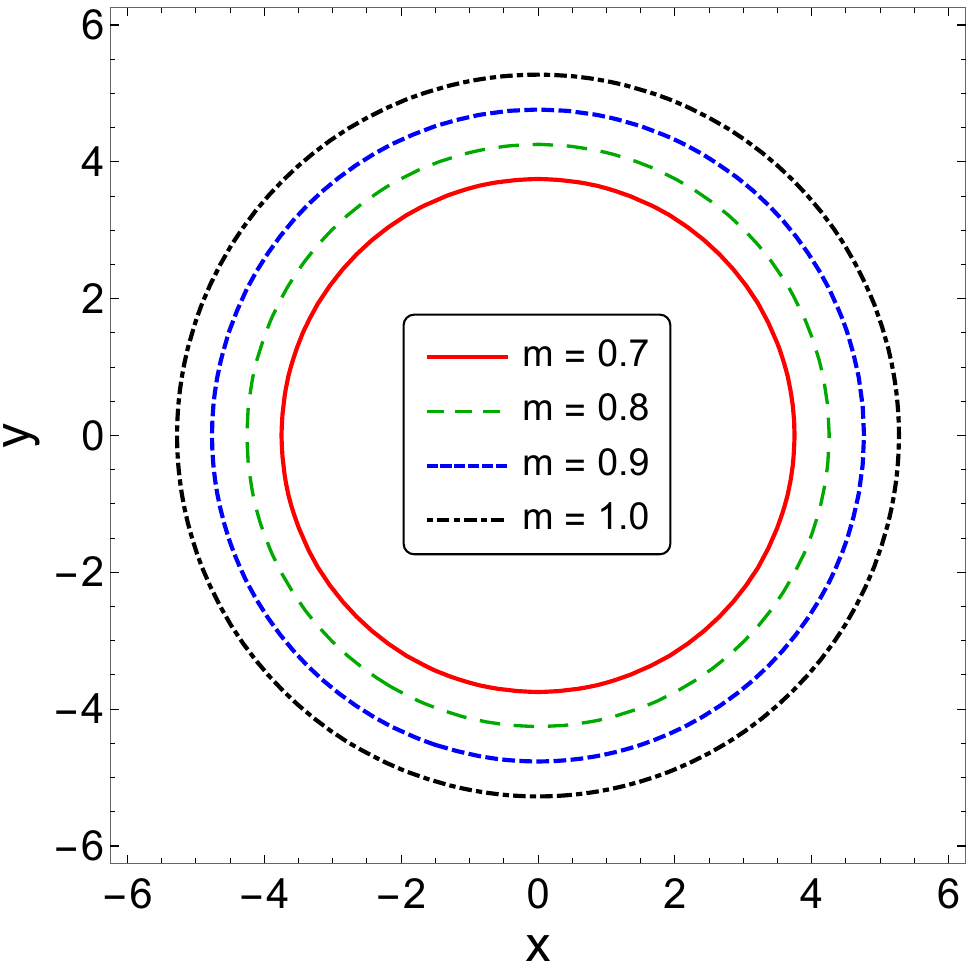}
	\end{center}
	\caption{The shadow of the Schwarzschild BH in NC spacetime for different values of the NC parameter $\Theta$ (left panel) with $m=1$, and for varying BH mass $m$ (right panel) with $\Theta=0.4$.}\label{fig2.7}
\end{figure}

Fig.~\ref{fig2.7} illustrates the shadow of the NC Schwarzschild BH for different values of the NC parameter $\Theta$ (left panel) and the BH mass $m$ (right panel). As shown in the left panel, the shadow radius increases with increasing $\Theta$. Similarly, the right panel indicates that for a fixed NC parameter, the shadow radius grows as the BH mass $m$ increases. This behavior suggests that the effects of non-commutativity and BH mass on the shadow radius are analogous. The NC parameter influences the spacetime geometry in a manner that mimics an increase in mass, effectively strengthening the gravitational field. This conclusion is consistent with our earlier results \cite{abdellah1}. Furthermore, we observe that the shadow radius of the NC Schwarzschild BH is larger and exhibits a slight increase with increasing $\Theta$, in contrast to the findings in \cite{Hassanabadi2}, where a significant increase with increasing $\Theta$ was reported.

Armed with the above result and considering the constraint on the NC parameter namely, that the NC correction must be smaller than the precision of the experimental measurement we choose to implement the NC shadow radius of the Schwarzschild BH. Specifically, we consider a typical primordial BH at the final stage of inflation with a mass of $GM \sim 5 \times 10^{-5} m$ \cite{karimabadi1} and a volume of $r \sim 1.5 \times 10^{-4} m$. Moreover, the precision of the experimental measurement, according to \cite{ehc}, is approximately $\sim 0.1$. Consequently, the upper limit of the NC parameter can be determined as:
\begin{equation}
	\Bigg|\frac{R_{\text{shadow}}-\hat{R}_{\text{shadow}}}{R_{\text{shadow}}}\Bigg|= \frac{(25\sqrt{3}-2)}{432m^2}\Theta^2\leq0.1\,.
\end{equation}

According to Ref.~\cite{PLANCK1}, the physical distance measured at any time after inflation can be obtained by multiplying our results by the square of the cosmological scale factor, $a = 10^{-29}$. This follows from our choice of the space-space NC matrix. By applying the above constraint to our system, we find that the NC parameter satisfies:
\begin{equation}
	\Theta^{\text{Phy}} = \sqrt{a^2\Theta^2} \leq 5.11 \times 10^{-34}\,\text{m}.
\end{equation}

It is evident that the obtained result is close to the Planck scale and is consistent with the bounds derived in our previous works, where similar constraints were obtained through experimental tests of gravity in the context of the NC gauge theory of gravity \cite{abdellah1,abdellah4} or by employing Bopp's shift transformation \cite{larbi1}, and even for the NC geometry in flat spacetime \cite{mirza,rome1}. These studies suggest a bound in the range of $(10^{-34}-10^{-31})$ m.

Moreover, we can obtain another constraint on the NC parameter using the unstable photon sphere, based on the same system described above. In this case, the lower bound of the NC parameter, derived from the unstable photon sphere with the accuracy of the Event Horizon Telescope \cite{ehc}, is given by:
\begin{equation}
	\Bigg|\frac{r^{\text{uns}}_{\text{ps}}-\hat{r}^{\text{uns}}_{\text{ps}}}{r^{\text{uns}}_{\text{ps}}}\Bigg|= \left(\frac{15 + 74 \sqrt{3}}{648m^2}\right)\Theta^2\leq0.1\,.
\end{equation}
and that lead to the new constrain:
\begin{equation}
	\Theta^{\text{Phy}}\leq 3.36\times 10^{-34}\,\text{m}.
\end{equation}
As we can see, this constraint is of the same order as the one obtained from the shadow radius of this BH. The difference arises due to the unstable photon sphere being closer to the NC event horizon compared to the shadow radius. 

Furthermore, it is possible to generalize this application to examine the constraint on the NC parameter from the innermost stable circular orbit (ISCO) of a massive test particle and to compare it with the constraint obtained from the photon orbit. The outer ISCO can be determined using the effective potential with $h = 1$ and $E = c = 1$, together with $l_{crit} = 3.5 m$. By solving the following condition, $\frac{dV_{\text{eff}}}{dr} = 0$, we obtain:
\begin{equation}\label{eq:4.58}
	\hat{r}^{\text{ISCO}}_{\text{outer}} \sim 7m + \left(\frac{0.0677306}{m}\right)\Theta^2\sim 7m + \left(\frac{\sqrt{17}}{61m}\right)\Theta^2\,.
\end{equation}

We assume that the observation of any close orbit of a massive or massless test particle is subject to the same accuracy constraint, which is approximately $10\%$ \cite{ehc}. Consequently, the constraint on $\Theta$ from the ISCO of a massive particle is given by:
\begin{equation}
	\Bigg|\frac{r^{\text{ISCO}}-\hat{r}^{\text{ISCO}}_{\text{outer}}}{r^{\text{ISCO}}}\Bigg|= \left(\frac{\sqrt{17}}{427m^2}\right)\Theta^2\leq0.1\,.
\end{equation}
which give us:
\begin{equation}
	\Theta^{\text{Phy}}\leq 1.61\times 10^{-33}\,\text{m}.
\end{equation}

It is clear that the constraint obtained from the trajectory of a massive particle is larger than that obtained from the photon trajectory, which aligns with our previous findings in Ref.~\cite{abdellah4}. Notably, the inner ISCO and the stable photon sphere are unique predictions of NC spacetime and can only be determined numerically. Any constraint derived from these new orbits requires direct observational data to obtain a numerical resolution that allows for a meaningful comparison with experimental observations. We believe that using these two new circular orbits provides a better estimation of the NC parameter compared to the outer orbits. 

It is particularly noteworthy that as we move closer to the event horizon of the BH, the NC parameter is constrained to smaller values. As demonstrated in the previous comparison between the unstable photon sphere and the shadow radius, where $\hat{r}^{\text{uns}}_{\text{ps}} < \hat{R}_{\text{shadow}}$ leads to a smaller value of $\Theta$. Thus, since the new circular orbits are closer to the event horizon, the constraint on the NC parameter derived from the inner ISCO or the stable photon sphere is expected to be smaller than those obtained in previous discussions, bringing it even closer to the Planck scale.

Remarkably, in this work, the estimation of the NC parameter is directly related to the accuracy of the experimental measurement of the BH shadow radius. This implies that as the precision of the measurement improves, the estimation of the NC parameter becomes more refined. A similar observation was made in Ref.~\cite{abdellah4}, where variations in experimental accuracy led to different estimations of the NC parameter.

\section{Conclusions}

In this study, we examined the geodesic motion of both massive and massless test particles in the NC Schwarzschild spacetime. Our approach was based on modifying the Schwarzschild BH metric using the NC gauge theory of gravity \cite{abdellah1}. In this geometry, we show that $\hat{g}_{tt}\hat{g}_{rr} \neq1$ is a necessary condition to satisfy the four energy conditions. This implies that the BH exhibits a double-photon-sphere structure: the inner photon sphere, which is stable, and the outer one, which is unstable. We computed corrections to the effective potential up to second order in \(\Theta\) and obtained modifications to both the affine parameter and the coordinate time in the NC Schwarzschild BH. Our results indicate that due to non-commutativity, both massive and massless particles require an infinite amount of time to reach either the NC singularity or the NC event horizon.

Furthermore, we analyzed the stability of geodesic motion using the Lyapunov exponent to assess circular orbit stability for both massive and massless particles. For massive particles, we determined the angular frequency \(\hat{\Omega}_c\), the time period \(\hat{T}_c\), and both the proper-time and coordinate-time Lyapunov exponents, \(\hat{\lambda}_p\) and \(\hat{\lambda}_c\), respectively. Notably, the proper-time Lyapunov exponent revealed a stable orbital region between two unstable regions near the event horizon, aligning with previous findings \cite{abdellah1}. For massless particles, our analysis of the effective potential uncovered a novel type of motion near the event horizon, indicating the presence of a stable circular orbit. As a result, a second photon sphere emerges in NC spacetime. Consequently, the NC Schwarzschild BH features two distinct photon spheres: a stable inner photon sphere and an unstable outer photon sphere. Using the coordinate-time Lyapunov exponent \(\hat{\lambda}_c^{\text{Null}}\), we confirmed the instability of the outer photon sphere, while the inner photon sphere remains stable— a feature unique to NC spacetime. Additionally, we computed the critical exponent \(\hat{\gamma}^{\text{Null}}\) to further assess the instability of null circular orbits in the NC Schwarzschild BH.

Lastly, we explored the impact of non-commutativity on the BH shadow. Our results suggest that non-commutativity strengthens the BH's gravitational field, mimicking the effect of increased mass. Key insights from our analysis include the connection between newly identified stable circular orbits near the event horizon and non-commutative effects, as well as an estimated upper bound on the NC parameter through the shadow radius, unstable photon sphere, and outer ISCO for massive test particles, which is approximately \(\Theta^{\text{Phy}}\sim (10^{-33}-10^{-34})\) m. Furthermore, our results highlight the impact of this geometry on orbital stability and geodesic motion around deformed compact objects (such as BHs), predicting new types of motion near the event horizon and introducing a new fundamental length scale, supporting the emergence of quantum effects at macroscopic levels. The study of other astronomical phenomena could further enhance our understanding of quantum gravity effects on large-scale systems.

\acknowledgments

This work is supported by project B00L02UN050120230003, Univ. Batna 1, Algeria.

\bibliographystyle{unsrt}
\bibliography{biblio2}
\end{document}